# Dynamic inhomogeneity, pairing and superconductivity in cuprates.

DRAGAN MIHAILOVIC AND VIKTOR V. KABANOV

*Jozef Stefan Institute and Dept. of Mathematics and Physics, Jamova 39, SI-1000 Ljubljana, Slovenia*

Dragan.mihailovic@ijs.si

**Abstract**

In cuprates there is a significant body of evidence for the existence of electronically driven dynamic inhomogeneity which might arise from the existence of polarons, bipolarons or other charged objects such as stripes, whose presence may be an essential ingredient for high-temperature superconductivity. In this review we examine the experimental evidence for such objects, defining the length, time and energy scales of the relevant elementary excitations. The dynamics of the objects below and above $T_c$ are examined in detail with femtosecond spectroscopy and compared with magnetic and other measurements. The dynamically inhomogeneous state is described theoretically by considering an interaction between electrons, spins and the lattice. By symmetry, only electrons in degenerate states can couple to the lattice and spins to give an *anisotropic*, *d*-wave symmetry interaction. The proposed interaction acts on a mesoscopic length-scale, taking into account the interplay of Coulomb repulsion between particles and anisotropic elastic strain, and leads to the formation of bipolaron pairs and stripes. The predicted symmetry breaking associated with pairing and stripe formation are observed in numerous experiments. The phenomenology associated with the co-existence of pairs and clusters (stripes) is found to apply to many different experiments ranging from femtosecond dynamics to transport measurements. Excitations of the system are described quite well in terms of a 2-level system, although we find that a complete description may require a more complicated energy landscape due to presence of mesoscopic objects such as stripes or clusters. The formation of the superconducting state can be understood quantitatively to be the result of the establishment of phase coherence percolation across pairs and stripes.

**Key words** *superconductivity   inhomogeneity      femtosecond spectroscopy      Jahn-Teller effect     local pairing   stripes*

## Introduction

Inhomogeneity appears to play an important role in determining the fundamental physics of many oxides, including superconducting cuprates, complex manganites and relaxor ferroelectrics. Even though the origins of the observed inhomogeneity are not yet entirely understood, it is believed to be related to nano-scale charge and spin ordering, driven by interactions between the doped charge carriers and their interaction with the lattice. Competition between interactions, particularly Coulomb and lattice deformations on different length-scales may lead to meso-scale ordering of charges and spins, reflected by phonon anomalies at finite wavevectors, phase separation, nano-scale pattern formation and incommensurate spin nanostructures.

Signatures of inhomogeneity have been found by very different experiments[1], probing on various time- and length-scales - in real space or in *k*-space – often leading to seemingly different and inconsistent interpretations. Many techniques typically give information which involves either spatial or temporal averaging (or most often both) and thus reveal different aspects of the problem from their respective viewpoints. Many such experiments which probe average lattice or electronic properties fail to observe inhomogeneities and some of these apparent discrepancies are not yet clearly understood. For example, local probe techniques which freeze the motion of ions on short timescales (e.g. XAFS[2] or neutron PDF studies[3]) which are eminently suitable for detection of these meso-scale patterns and deviations of lattice structure from the average, appear to be inconsistent with slower techniques such as NMR[4], or standard diffraction techniques which also appear to show no evidence of static inhomogeneity. For example, recent resonant X-ray scattering experiments have not shown any evidence of charge ordering or inhomogeneity. The resolution in these experiments, however, is limited by the X-ray wavelength of $\lambda = 2.2$ nm, and the features on a length-scale of the coherence length $\xi_s$ ~1-2 nm might be missed[5]. A different problem occurs with optical experiments, such as femtosecond pump-probe experiments, which have spatial resolution which is limited by the wavelength of light (typically 800 nm).

Yet, a multi-component response, as a sign of an inhomogeneous state has been consistently observed in time-resolved and many other experiments, such as ESR[6], NMR[7], magnetic susceptibility[8], as well as optical conductivity[9] and ARPES[10] (peak-dip-hump structure).

One possible way to resolve these discrepancies is to assume that inhomogeneity dynamics occurs on a timescale which is relatively fast. Although this cannot explain the static surface inhomogeneity observed in scanning tunnelling microscope measurements[11], which show both a superconducting gap and pseudogap structure on the surface, these static features might be due to surface pinning on defects or impurities.

Thus, the experimental challenge of the last decade has been to invent and perfect new techniques for the investigation of bulk dynamic inhomogeneities on time-scales sufficiently short to give information on the dynamics of the relevant excitations giving new insight into the microscopic origins of the dominant interactions leading to the observed dynamic complexity.

The experimental observations of inhomogeneity have also posed fundamental challenges for theories attempting to describe the multi-component inhomogeneous ground state in a consistent and unified manner - and leads to a description of superconductivity. The interplay of different phases with different symmetry, interacting on a mesoscopic scale is thought to be particularly interesting, and will be explored in some detail.

The review is divided into four parts. First we briefly summarise some of the evidence and arguments for the existence of lattice inhomogeneity and the inhomogeneity of the electronic structure. In the second part, we concentrate on a series of time-resolved optical experiments, which give a great deal of information on charge carrier dynamics on fast timescales, revealing ubiquitous presence of multi-component pair recombination dynamics associated with the presence of a dynamic spatially inhomogeneous electronic structure. The distinguishing feature of time-resolved experiments is that they can distinguish between different excitations even when these have the same energy scale, i.e. situations where spectroscopies such as ARPES, Raman, infrared, STS etc. - which cannot intrinsically distinguish between homogeneous and inhomogeneous lineshapes - fail. The important and generic results of time-resolved experiments on cuprates are analysed theoretically using a phenomenological level using effectively 2-level systems, which leads to a 2-component description of the ground state. The inhomogeneity is inferred to be mesoscopic or microscopic rather than macroscopic.

In the third section, our theoretical understanding of the electronic ground state and pairing is developed. The line of reasoning which we use is motivated by the original ideas of G.Bednorz and K.A.Muller on Jahn-Teller polarons, which lead to the discovery of high-temperature superconductivity in $La_{2-x}Ba_XCuO_4$[12]. A model Hamiltonian is proposed to describe the symmetry-allowed interactions which are consistent not only with the time-resolved experiments, but also numerous others, such as NMR, bulk susceptibility, Raman and ARPES etc. Some of the important and very specific manifestations of the proposed model interaction are discussed, particularly related to spatial symmetry breaking, in agreement with observations of such phenomena in the superconducting cuprates.

Finally in the last section we consider how the proposed picture can lead to superconductivity and describe the phase diagram arising from the inhomogeneous-state description. In each section, the issues are separately justified with the interpretations as concise as possible, and limited to the issues in hand. We also include criticisms and open issues wherever these are thought to be important.

## Evidence for inhomogeneity: length, time and energy scales

At present there is no single technique which can reliably give accurate spatially-resolved, real-time picture of the nano-scale lattice dynamics, electronic or spin structure, and these have to be deduced by combining the results of different techniques. For example, local probe techniques such as XAFS and neutron scattering with pair-distribution function (PDF) analysis, give information on near-instantaneous positions of atoms in relation to their nearest neighbors, but their accuracy is limited by the limited range of wavevectors $k$ which can be measured, and have pitfalls in modeling the data, leading to ambiguity in atomic displacements which are deduced[2]. NMR has an even shorter range, and has a relatively slow timescale of the order of $10^{-7}$ s, averaging out any dynamics which might be occurring on faster timescales. STM is limited to the surface, and is a quasi-static probe of the inhomogeneity in the density of surface states, but has nevertheless received a great deal of attention lately.

In the time-domain, optical time-resolved techniques can accurately measure electronic relaxation times and pair recombination processes in the bulk on timescales from femtoseconds to microseconds, but since the spatial resolution is limited, the existence of spatial inhomogeneity is deduced on the basis of the observation of multi-component relaxation processes and modelling of the electronic structure.

To obtain an overall coherent picture we will discuss the length, time and energy scales respectively, of inhomogeneities inferred from different experiments in the next section.

### *Length-scales: real space*

Direct measurements of a dynamical inhomogeneities in bulk are not yet within the realm of experimental possibility, so information on length-scales needs to be pieced together from different techniques: Scanning-tunnelling microscopy (STM) experiments of DeLozanne et al [11] on the surface of $YBa_2Cu_3O_{7-\delta}$ and more recently by the Berkeley groups on $Bi_2Sr_2CaCu_2O_8$ [11] show the local density of electronic states at different points on the surface to be very inhomogeneous, with a characteristic length-scale of the order of $l_0 \sim$ 1-2 nm. The surface inhomogeneity clearly shows the lattice symmetry properties, e.g. tetragonal *d*-wave-like symmetry surrounding single defects for example[11]. Certainly BiSCO and YBCO superconductors appear to be inhomogeneous on the surface, yet the two materials differ significantly in the details. YBCO shows quite clear stripe-like features along the crystallographic *b* axis - which appear to be connected with its Cu-O chain structure - while BiSCO shows features with typically 4-fold symmetry. Unfortunately the exposed surfaces observed by STM are *not* $CuO_2$ layers in either case, but intermediate layers in the charge reservoirs (Bi-O layers or Cu-O chains respectively). Whether these inhomogeneities are also present in the bulk cannot be deduced from STM experiments, but the observed length scale is clearly in the range of $l_0 \sim$ 1-2 nm, with some additional hints of the existence of larger objects, resembling charge or spin density wave segments, observed by Howald et al[11].

A structural length scale can also be deduced from diffuse diffraction measurements. For example, Kimura et al[13] deduced a structural coherence length of 1.7 nm in underdoped $La_{1.9}Sr_{0.1}CuO_4$ at 315 K in (i.e. near the pseudogap temperature). The structural coherence length is deduced from the *k*-linewidth of the diffuse central peak ($\Delta k$=0.06 Å$^{-1}$). Well below $T_c$ (at T=13.5 K) the structural coherence length is extended to $l_s >$ 10 nm (the linewidth is resolution limited). Unfortunately there are presently no detailed studies of structural inhomogeneities in LSCO over a large range of temperatures, but recent studies on superconducting YBa$_2$Cu$_3$O$_{6+x}$ have revealed the existence of c-axis order as well as in-plane inhomogeneity on a similar length-scale of 1-2 nm above T$_c$. Islam et al[14] find diffuse peaks in neutron scattering which originate from mutually coupled, primarily longitudinal displacements along the *a* axis of the CuO$_2$, CuO chain and apical O or Ba atoms. The characteristic *T*-dependence of these peaks shows them to be related to the pseudogap $T^*$. A similar temperature dependence is also displayed by atomic displacements observed in ion-channeling experiments[15].

The structural coherence length scale appears to be linked with the length scales deduced from spin-ordering[16], which also shows similar temperature dependence.

We can also obtain an intrinsic structural coherence length scale from the QP lifetimes measured by femtosecond spectroscopy. The QP recombination time $\tau_R$ is related to the anharmonic lifetime $\tau_{anh}$ of high-frequency phonons involved in the QP recombination process, whose lifetime in turn is limited by the escape time of their decay products, namely acoustic phonons. Thus $\tau_R$ is essentially determined by the acoustic phonons' mean free path, which in determined by the length-scale of structural inhomogeneities. So the structural length scale $l_0 = v_s \tau_R$ where $v_s$ is the sound velocity. From the data on $\tau_R$ for YBCO and other cuprates, near $T_c$, we obtain - rather remarkably - a characteristic length of $l_0 \approx$ 2 nm. $l_0$ decreases slowly above $T_c$ but increases more rapidly below $T_c$. The length scale determined from the recombination time will be discussed in more detail later in the section on QP dynamics.

The fact that the superconducting coherence length in hole-doped cuprates $\xi_s \sim 1-2$ nm is very similar to the inhomogeneity length scale $l_0$ at temperatures near the superconducting $T_c$ is probably not a coincidence, and has important implications for superconductivity, and particularly supporting the existence of percolative superconductivity.

### *k-space –derived length scale*

Inelastic neutron scattering experiments on phonons in hole-doped $YBa_2Cu_3O_{7-\delta}$[17], $La_{2-x}Sr_xCuO_4$ [18] and inelastic X-ray scattering in electron-doped $Nd_{1.86}Ce_{0.14}CuO_{4+\delta}$ [19] all show a strong phonon anomaly starting close to the middle of the Brillouin zone (BZ). The data show a high-frequency optical phonon mode near 85 meV, which is well-defined at the zone centre, but becomes strongly damped and virtually dissappears in the middle of the BZ. A mode then re-appears at near 70 meV towards the edge of the BZ in the $(\zeta,0,0)$ direction (corresponding to the M point in LSCO). The wavevector of the center of the anomaly $k_0$ corresponds rather closely to the length-scale $l_0$ of real-space textures in STM images as discussed in the previous section. The two can be related by $k_0 \sim \pi/l_0$, where $\pi/k_0 = 2 - 3$ unit cells (0.8-1.2 nm) in LSCO[18]. The anomaly appears over a relatively range of wavevectors $\Delta k$ corresponding to the range of spatial distortion with a range of $\Delta l_0$.

The fact that $\Delta k$ spans approximately half the BZ, implies that there is no long range order associated with these anomalies, which would also imply the existence of clear new peaks in diffraction experiments. Long range ordered charge stripes would appear over a much narrower range of $k$ than the nano-scale objects seen by STM, and their existence can also be excluded by the STM, INS data discussed above, as well as resonant X-ray scattering[5] and many other diffraction techniques. Moreover commensurate or incommensurate *static* stripes should give rise to a clearly observable zone folding, but this is not observed in X-ray diffraction (XRD) or Raman scattering, supporting the view that the inhomogeneous structure is dynamic. We conclude that the INS and inelastic X ray scattering experiments speak for the existence of dynamic bulk inhomogeneities which have a similar length-scale $l_0$ as the ones observed on the surface by STM.

It should be mentioned that such anomalies in *k*-space have also been observed in other non-superconducting materials such as $La_2NiO_4$. The details (e.g. temperature-dependence, characteristic range $\Delta k$, and general texture of the nanoscale structure) are different however. Indeed a concise experimental comparison with superconducting cuprates still needs to be performed. The existence of phonon anomalies in $La_2NiO_4$ and $La_2MnO_4$ implies the existence of charge inhomogeneity and also charge ordering, but does not necessarily imply superconductivity as well. The view of the present authors is that the occurrence of superconductivity depends on whether the inhomogeneity is associated with pairs, single polarons or a charge-ordered (CDW) state. As we shall see later, the evidence for spin singlets[20,48] and absence of a Curie susceptibility in the cuprates overwhelmingly supports the view that the inhomogeneity is associated with singlet pair formation.

### *Time-scales: Dynamic probes*
Possibly the first experiments which unambiguously showed the presence of an inhomogeneous *electronic* structure of bulk materials on a picosecond timescale were non-equilibrium Raman experiments with picosecond laser pulses, which clearly showed the presence of localized states in metallic, optimally doped YBCO[21]. The activation energies for hopping of photoexcited carriers was found to be close to the „pseudogap energy" scale $E_p$ in YBCO, with $E_a = 34$~$210$ meV, depending on doping.

The experiments also showed clearly that the hopping process was coupled to the lattice[22]. Subsequent more detailed and systematic time-resolved pump-probe experiments (to be explained in more detail in the following section) showed the presence of multiple intrinsic relaxation times, which appeared to confirm the co-existence of localised and itinerant states[23,24] with vastly different timescales. The pair recombination dynamics appears on a timescale of $10^{-13}$ s, while localised states appear to have a distribution of lifetimes which extend to hundreds of microseconds[24,25].

XAFS and neutron PDF experiments probe the local structure on time-scales of the order of $10^{-15}$-$10^{-12}$ s. In the case of XAFS, the motion of excitations whose energy scales are less than ~500 meV is effectively „frozen". This energy scale is a few tens of meV in the case of neutron PDF. Significantly, XAFS structure snapshots are on a timescale, which is faster than both the pair recombination time and lattice motion. This means that it can be used to „freeze" the structure of lattice distortions associated with local pairs, which occur on an energy scale of the pseudogap (< 100 meV), hence the techniques have been very important in pointing out the presence of inhomogeneities in cuprates and other oxides[26]. The interpretation of these structures in terms of long-range ordered stripes has proved controversial[2], although there appears to be an emerging consensus that significant deviations from an average structure exist on timescales $10^{-10}$ to $10^{-14}$ seconds.

Another clear indication of the relevant timescale comes from the linewidths of anomalous phonon spectra observed in INS around $k_0$, which are of the order of $\Delta E \sim$ 4 -5 meV[18]. The implied lifetime of the excitations is $\tau = h/\pi c\tau \sim 300$ fs. This is comparable with the superconducting pair recombination lifetime of $\tau_R \sim 300 - 1000$ fs as measured directly by femtosecond optical techniques using time-resolved pump-probe excited state absorption[23]. A similar lifetime is obtained by THz radiation pump-probe[27] experiments which directly measure the condensate recovery as a function of time after excitation by 70 femtosecond optical pulses. The connection between the appearance of charge-related inhomogeneity in INS and pair recombination measured by optical time-resolved techniques is clearly implied.

***The energy scale of the relevant interactions***
The energy scale of the anomalies in INS and ARPES is ~50-100 meV, and is the same as the magnitude of the pseudogap $E_p$ in the equivalent doping range, as determined by single particle tunneling[28], femtosecond timescale pump-probe recombination experiments[23,29,30], non-equilibrium Raman experiments[22], and other techniques which measure the *charge* excitation spectrum. As an example, in Figure 1, we compare the "pseudogap" magnitude determined by QP recombination[23] and tunneling[28] with the anomaly observed neutron data[17] (shaded area). The excitation has the same energy scale, virtually the same lifetime and a similar temperature dependence, confirming that the „object" observed in INS and time-resolved experiments is one and the same. The energy scale of inhomogeneity thus appears to be defined by the charge-excitation „pseudogap" $E_p$. From the observation that *charge* excitations in YBaCuO typically show a higher energy gap than *spin* excitations[20], it would appear that spin inhomogeneity and ordering follows the formation charge inhomogeneities, which seemingly occur first.

Another energy scale in cuprates is defined by the superconducting $T_c$, but at present we are not aware of any evidence of a link between the appearance of inhomogeneities and the onset of superconductivity. Thus inhomogeneity appears to be intimately related primarily to the „pseudogap" behaviour in cuprates. This confirms that superconductivity itself is associated with the Γ point of the Brillouin zone, i.e. the zero-momentum state.

## Time resolved experiments probing the dynamics of the inhomogeneous state.

The measurement of the time-resolved optical response in general gives detailed information on low-energy excited state lifetimes, more specifically on quasiparticle recombination and related inhomogeneity dynamics. Generally, the pump-probe technique involves a measurement of the change of reflectivity (or transmittance, for the case of a thin film) of the superconductor as a function of time after a short „pump" laser pulse excitation. Although the energy of laser photons used to excited the superconductor is large compared to the gap, the photoexcited charge carriers quickly relax from 1.5 eV to low energies by electron-electron and electron-phonon scattering. This process is known to proceed on a timescale of ~ 10-100 fs. A gap in the low-energy excitation spectrum (such as occurs in a superconductor or semiconductor or charge-density-wave system), may lead to a bottleneck situation where QPs accumulate at the bottom of the band (see Figure 2). These photoexcited quasiparticles are detected by the probe photons using a suitably delayed laser pulse by what is effectively an excited state absorption of photoexcited quasiparticles (as shown in Figure 2) [29,30]. The measurement of the amplitude of the changes in probe pulse reflectivity (or transmittance) as a function of time directly measures the recombination dynamics. The temperature dependence of the excited state absorption in pump-probe measurements can be used to determine the temperature dependence of the gap, its magnitude, as well as - to some extent - its symmetry on pairing timescales.

Following initial pioneering experiments of time-resolved reflectivity on cuprates[31], the technique has been shown to give excellent reproducibility from sample to sample, and data from different groups have been in close agreement. (The variances which have been reported so far have been related to the low-temperature lifetime (well below $T_c$) which appears to be strongly dependent on the heating caused by the pump laser pulses, and possibly also sample quality[32].)

***Experimental observations of multi-component relaxation.***
The typical time-resolved reflectivity response for YBCO is shown in Figure 3. The response at short times (from 0.1 to 5 ps) is qualitatively the same in all HTS materials. At short times after photoexcitation a fast response is observed with $\tau_1$ = 0.3 ~ 1 ps (depending on material), which is almost *T*-independent, followed at longer times (but still on the picosecond timescale) by a somewhat slower decay. This second decay appears only *below* $T_c$, and its lifetime is *T*-dependent, showing clear signs of divergence as $T_c$ is approached from below. (The divergence of $\tau_S$ is unambigous indication that it represents recombination across a superconducting gap which is temperature dependent near $T_c$, since $\tau_S \propto 1/\Delta_s(T)$, and $\Delta_s(T) \to 0$ as $T \to T_c$.).

In addition to the response on the 0.1-3 ps timescale, a ubiquitous long-lived response is observed which has been attributed to localised states[24]. This will be discussed later.

The two fast responses may have the same or opposite sign in different materials, which is a consequence of the different optical processes of the probe pulse[29] (shown in Fig. 2). In YBCO[23], LaSCO, NdCeCuO[33] and HgBaCaCuO[29] they typically have the same sign for 800 nm probe wavelength (YBCO and Hg-1223 are also shown in Fig.3), while in Tl-based and Bi-based cuprates they have opposite signs[34]. Remarkably, the temperature dependence of the amplitude (i.e. magnitude of photoinduced reflectivity or absorption) of the two responses appears to be quite universal in the cuprates (Figure 4). A careful analysis of the amplitude of the two signals gives two distinct temperature dependences. The superconducting state response $\tau_s$, always disappears abruptly at $T_c$, while the other faster response $\tau_p$, disappears asymptotically at much higher temperature $T^*$. The component with the fast response, whose relaxation time is $\tau_p$, can be unequivocally associated with the „pseudogap", while the slower response with a lifetime $\tau_s$ is associated with the appearance of a superconducting gap.

In the Hg-cuprate and YBCO 124, where the two signals have the same sign and the lifetimes $\tau_s \approx \tau_p$ the deconvolution of the two signals is more ambiguous. Nevertheless a comparison with YBCO, LSCO, Tl and Bi- based superconductors shows quite convincingly that the universal 2-component dynamics on the picosecond timescale and 2-component temperature-dependence is observed in all cuprates investigated so far. This universal multi-component response is clear indication of the ubiquitous presence of inhomogeneity in cuparates.

### *Modeling in terms of 2-level systems*

The description of quasiparticle recombination across a supeconducting gap for classical superconductors was first considered by Rothwarf and Taylor[35], who postulated a set of differential equations, describing QP recombination via the emission (and re-absorption) of phonons. In classical superconductors the QP recombination lifetime is typically $\tau_R \sim 10^{-9}$ s. In HTS materials, the gap is quite large and is comparable to high-energy phonon frequencies, $\Delta \sim \hbar\omega_p$, and the recombination timescale is in the picosecond range[29].

The recombination dynamics in HTS was considered in detail theoretically by Kabanov et al [29]. In addition to the calculation of the lifetime (which follows the approach by Rothwarf and Taylor[35]), they also calculate and verify the temperature-dependence of the QP population under bottleneck (i.e. near steady-state) conditions. Using an effectively 2-level system description, expressions for the temperature-dependence of the QP population - and hence of the T-dependence of the transient optical response - were derived.

Two cases are considered, (i) a *T*-independent gap (applicable to the formation of local bipolarons for example, which can be described by a 2-level system) and (ii) a *T*-dependent superconducting gap (e.g. a BCS-like $\Delta(T)$) (shown schematically in Figure 5). The *T*-dependence of the QP density for the two cases is given by two expressions:

$$n_{QP}(T,\Delta_p) \propto \frac{1}{\Delta_0}\left\{1+\frac{2\upsilon}{N(0)\hbar\Omega}e^{-\Delta_0/k_BT}\right\}^{-1}, \qquad \text{for } \Delta_p = \text{constant} \qquad (1)$$

and

$$n_{QP}(T,\Delta_{BCS}) \propto \frac{1}{\Delta(T)+k_BT/2}\left\{1+\frac{2\upsilon}{N(0)\hbar\Omega}\sqrt{\frac{2k_BT}{\pi\Delta(T)}}e^{-\Delta(T)/k_BT}\right\}^{-1}, \qquad \text{for a BCS-like }\Delta(T). \quad (2)$$

where N(0) is the density of states, $\nu$ is the number of phonons emitted, $\Omega$ is the phonon frequency[29].

In case the gap $\Delta$ is temperature *independent*, the response is predicted to fall off asymptotically at high temperatures. In the case where $\Delta$ is *T*-dependent, and closes at $T_c$, the signal must also fall to zero at $T_c$. Both *s*-wave and *d*-wave gap cases have been calculated..

Comparison of the theory with experimental data of both the *T*-dependence of the non-equilibrium QP density (Figure 4) and lifetime (Fig. 6) gives remarkable agreement for each component separately. The pseudogap relaxation dynamics is modelled with a *T*-independent gap (Eq. 1), while the superconducting gap is modelled using a BCS-like gap $\Delta_s(T)$ (Eq. 2).

A fit to the data in Fig. 4 universally shows that the two components in the time-resolved response imply a *simultaneous* presence of a superconducting gap $\Delta_s(T)$ and a *T*-independent gap $\Delta_p$. The latter is interpreted as the bipolaron (pair) binding energy, i.e. the energy scale associated with the appearance of inhomogeneity.

This experimentally observed 2-component dynamics is highly unusual and cannot be easily understood. For example, it cannot be described as a cascade of processes in a homogeneous medium. This is best seen from the data on $YBa_2Cu_3O_{7-\delta}$: the magnitudes of the two gaps $\Delta_p$ and $\Delta_s(T=0)$ are approximately equal in magnitude near optimal doping (Figure 7), which clearly rules out a relaxation cascade processes. The only way to describe the situation is to assume that the system is *inhomogeneous*, with the two different relaxation processes occurring in different areas (on the nanoscale), or different regions in *k*-space (with no cross-relaxation on the pairing timescale).

Remarkably, an excellent fit to the *T*-dependence of the optical response is obtained when an *s*-wave gap is used in expressions (1) and (2), for either component. On the other hand, Kabanov et al have shown that all attempts to fit the data with a substantially anisotropic gap or a simple *d*-wave gap, particularly for $\Delta_s(T)$ fail[29]. Both the *T*-dependence and the intensity dependence are distinctly s-like. This discrepancy between the predicted behaviour for a simple *d*-wave gap and the time-resolved data has proved to be very robust. Experiments from different groups have given similar data, none of which could be fit with a *d*-wave gap.

There are two possible ways to understand the apparent discrepancy. Either the model of Kabanov et al[29] is inappropriate, or our understanding of the gap structure on the femtosecond timescale is incomplete and the electronic structure of the cuprates based on simple *s* and *d*-wave gaps needs to be re-examined.

Fortunately, testing Kabanov's model has proved to be straightforward, because there exist a number of other materials with a gap in the low-energy spectrum, particularly charge-density wave (CDW) materials. Thus a number of quasi-1D and quasi-2D compounds were measured ($K_{0.3}MoO_3$, 1T-$TiS_2$ and 2H-$TaSe_2$ respectively) with time-resolved spectroscopy[36]. All showed excellent agreement with Kabanov's model, particularly the temperature (in)dependence of the gap(s) and relaxation time.

Moreover, systematic time-resolved studies as a function of doping in LaSCO and YBCO, as well as Hg-1223 and YBCO-124 have shown the *values* of the pseudogap $\Delta_p$ and superconducting gap $\Delta_s$ to be very close to those measured by single-particle tunneling or ARPES, and also follow the same systematic dependence on doping. (Fig. 7). The gap ratios $2\Delta/kT_c$ are systematically found to be in the range 8-10, for the materials measured so far, in agreement with numerous other techniques. We conclude that the gap structure on the pairing timescale cannot be described by a simple *d*-wave picture.

Kabanov et al [29] also calculated the *T*-dependence of the QP recombination lifetime in cuprates. The predicted lifetime is also dependent on the magnitude of the gap $\Delta$. In case $\Delta$ is temperature-dependent, then the lifetime shows a divergence of $\tau_s$ as as $\Delta \to 0$, as $T_c$ is approached from below:

$$\tau = \tau_{anh} \frac{\hbar^2 \omega_p^2}{12kT \Delta(T)}$$

Where $\tau_{anh}$ is the phonon anharmonic lifetime, $\omega_p$ is the characteristic frequency of the emitted phonon and $k$ is Boltzmann's constant. The lifetime at intermediate temperatures (at $T = T_c/2$) was estimated on the basis of experimentally observed anharmonic phonon lifetimes in YBCO[37], and was found to be of the order of 0.8 ps, in remarkable agreement with the observed optical transient lifetime of τ = 1 ps[29], considering all parameters were independently experimentally determined. More importantly, the predicted divergence has also been observed in many cuparates [29], below $T_c$ (such as in Figure 6).

The appearance of the divergent QP lifetime $\tau_s$ near $T_c$ is clearly consistent with the $T$-dependence of the QP density $n(T)$ and $\Delta_s(T)$ deduced from analysis of the $T$-dependence of the amplitude of the time-resolved signal below $T_c$ (Figure 4). In addition to the divergence at $T_c$ arising from $\Delta(T) \rightarrow 0$ as $T \rightarrow T_c$, the lifetime is expected to diverge at low $T$, as is indeed observed in many cuprates[29,32,34].

Another rather surprising fact is that when we use the value of the gap measured by pump-probe spectroscopy ($2\Delta_0$ = 900 K) at optimum doping[29], the gap ratio $2\Delta_0/kT_c$=10, which gives the recombination lifetime near $T_c$ as:

$$\tau = \tau_{anh} \frac{\hbar^2 \omega_p^2}{2.4\Delta_0^2}$$

Considering that $\hbar\omega_p \approx \Delta_0$, then the recombination lifetime is approximately equal to the zero-temperature anharmonic lifetime of the emitted phonon, $\tau \approx \tau_{anh}$.

The underlying physics of the recombination process was first discussed by Rothwarf and Taylor[35], where they considered the kinetics of the QP recombination in terms of the emission and re-absorption of phonons. As two QPs recombine, the energy is released to a phonon (there are no other excitations to carry off the energy), which has to have an energy $\hbar\omega_q > 2\Delta$, otherwise the recombination cannot take place (Figure 8). (In high-$T_c$ cuprate superconductors where the gap is large, we are limited to optical phonons). These optical phonons can also be efficiently re-absorbed in a pair-breaking process, thus creating a bottleneck in the QP relaxation process. The only way that the energy can be released is if the emitted optical phonon decays anharmonically to lower-energy phonons which carry off the energy in a way that it cannot be re-absorbed.

This process can be easily understood in the bipolaron pairing picture. Suppose two QPs recombine to form a bipolaron of size $l_0$. In this process, an optical phonon is emitted, but it remains in the Rothwarf-and-Taylor (RT) loop until the anharmonic decay products (i.e. acoustic phonons) leave the bipolaron volume. At that point, the pairs can no longer be excited. The most obvious consequence is that the lifetime is determined by the time that the acoustic phonons leave the bipolaron volume, which, to first approximation can be given by $\tau_R = v_s l_0$, where $v_s$ is the sound velocity (see Figure 9). (This process can also be treated more rigorously, taking into account the escape kinetics of acoustic phonons as a third kinetic equation in addition to the two RT equations[38], but the underlying physics is the same.)

The same picture holds for stripes, except that the $l_0$ describes the characteristic size of the stripe or cluster.

From the analysis presented above we see that we can determine the characteristic size of bipolarons or clusters from the measured QP lifetime $\tau_R$, by using the relation $l_0 = v_s \tau_R$. In Figure 10 we have plotted $l_0$ for $La_{2-x}Sr_xCuO_4$, Hg-1223 and $Nd_{2-x}Ce_xCuO_4$ as a function of normalised temperature, normalised to the superconducting coherence length $\xi_s$. Remarkably, in spite of the fact that $p$-type and $n$-type materials have very different coherence lengths, $l_0/\xi_s \approx 1$ near $T_c$. Moreover, the data exhibits approximate scaling behaviour $l_0 = \xi_s T^{-\nu}$ which might be expected for a quantum critical point (QCP) at $T = 0$, where the critical exponent of $\nu \approx 1$ is found typically in random systems, such as glasses. The fact that the observed behaviour is seen in all the cuprates (including $p$- and $n$-type materials) is a good indication that the underlying physics is quite universal.

### *k*-space description of QP recombination dynamics and gap anisotropy

Time-resolved experiments seem to suggest that there are very few, or no QP states at low energy. On the other hand, there are many experiments which *do* show a *d*-wave-like gap in the density of states, so we need to consider the possibility that the picture of the low-energy electronic structure at short times can be different than in the steady state. In other words, we need to take into account the short timescale of the experiments, and more specifically the *k*-dependence of the recombination processes involved in QP recombination on fast timescales.

First, let us consider the sequence of events after short laser pulse excitation. The initial excitation of electron-hole pairs by the laser pulse and the subsequent relaxation to states near the Fermi energy of these carriers takes place within a few tens of femtoseconds[39]. This process is reasonably well understood, and has been studied in normal metals, superconductors and semiconductors in some detail. Before reaching low energies, the carriers have undergone a cascade of scattering events, rapidly losing memory of their initial momentum, and occupying all available momentum states approximately equally (see Figure 11).

As already discussed in a real space picture (Figure 9), the next step in the energy relaxation process is the recombination of QPs to the ground state via the emission of one or more phonons[29]. If a phonon exists, whose energy is greater than the twice the gap, $\hbar\omega_q > 2\Delta$, then all the energy released in the QP recombination can be carried off in a first-order process by a single phonon. Such a process involves the recombination of two QPs with momenta $k_1$ and $k_2$, and energy $E_1$ and $E_2$, to a pair state with momentum $k_\text{pair}$ at $E_F$, and the emission of a phonon with momentum $q$ of energy $\hbar\omega_q$.

E and *k* conservation restrictions are quite different in the pseudogap state and in the superconducting state, because the final (pair) state momenta are different in the two cases (see Figure 12 a) and b)):

a) In the pseudogap state (Figure 12 a)), with $T^* > T > T_c$, pre-formed pairs may be mobile, and the final pair state momentum is determined by their center-of-mass kinetic energy, so there are no strict kinematic restrictions on the final state pair momentum, such as in the superconducting state. In describing the recombination process we can therefore simply consider energy conservation pertaining to a 2-level system of QPs in the excited states and bound pairs in the ground state, separated by a gap $\Delta_p$, [29]

b) Pairs in the superconducting condensate (Figure 12 b)), for $T < T_c$ are usually considered to have zero net momentum $k_{condensate} = 0$, so $\boldsymbol{k}_1 + \boldsymbol{k}_2 - \boldsymbol{q} = 0$, and the gap $\Delta_s$ separating QP states and the condensate pairs is T-dependent.

To obtain a better understanding of the recombination kinematics in this case, we have plotted the relevant processes in *k*-space Figure 12. To make the picture as realistic as possible, we discuss the recombination using an electronic structure based on the generic features observed in recent ARPES measurements on $La_{2-x}Sr_xCuO_4$ [40]. The spectral intensity $A(k,\omega)$ at 10 K obtained from the raw ARPES data on LaSrCO is shown schematically in Figure 12. Shades of gray represent the value of $A(k,\omega)$ at a binding energy $\Delta(k)$ corresponding to the peak in $A(k,\omega)$. These peaks - which are interpreted as a sign of the presence of QP states relevant in the recombination process - occur at an energy $E = E_F - E_\Delta$. $E_\Delta$ is doping-dependent and is shown plotted on the energy-vs.-doping diagram in Figure 7 as a function of doping *x*. The dashed arrows in Fig.12 b) represent the two types of QP recombination processes involving $k_{pair}=0$ pairs in the condensate as the final state, which satisfy momentum and energy conservation, $\boldsymbol{k}_1 + \boldsymbol{k}_2 - \boldsymbol{q}_R = 0$ and $E_1 + E_2 - \hbar\omega = 0$ respectively. It is easy to see that the emitted phonons (wavy lines) can have only certain values of momenta $q_R$ (shown shaded in Figure 12). At the same time they must have an energy $\hbar\omega_q > 2E_\Delta$ (60-80 meV), which in cuprates is limited to high-frequency optical modes involving predominantly O motion.

The conclusion from these analyses is that Kabanov's model for QP recombination is perfectly valid provided momentum selection rules are properly taken into account and realistic anisotropy of the band structure is taken into account. They also show that QP recombination proceeds primarily from the M points in the BZ. (i.e. the antinodes) with the highest DOS, while the experiments shown that there are virtually no recombination processes for QPs with momenta at the X point in the BZ (in the nodal directions). The pseudogap response clearly implies that the short-timescale picture with the main QP density at the M points can be described well by Kabanov's model. The superconducting state response on the other hand has stricter selection rules with the requirement that $\boldsymbol{q}_{\text{pairs}} = 0$, but again provided the real DOS is taken into account, Kabanov's model still describes the picture very well, of course considering each component separately.

In addition to the 2-component ultrafast response on the picosecond timescale discussed above, in all cuprates there is also a ubiquitous much slower response, with a timescale which is longer than 1 ns, which appears to be glass-like without a clearly defined lifetime[24,25]. This is present as in CDW systems as well[24], and is attributed to localised intra-gap states. It has been suggested that the origin of these processes may be related – at least in cuprates – to relaxation of nodal quasiparticles[41]. However, its appearance in many CDW systems such as $K_{0.3}MoO_3$ suggest that this model of d-wave nodal relaxation is probably not applicable, and that clearly this very slow relaxation component is not an exclusive feature of the *d*-wave state. Because these states are probably not very important for the mechanism of superconductivity, nor are they likely to *drive* the formation of an inhomogeneous state, we shall not discuss them further here.

### *Origin of the 2-component femtosecond response*

In summarising the time-resolved experiments, we infer two unusual features which are ubiquitous in the cuprates:

- The QP gap structure on the femtosecond timescale shows large gap behaviour, with negligible density of states at low energy, and dynamics which is similar to (more isotropically gapped) CDW systems.

- The recombination dynamics shows a 2-component response on the 0.2-5 picosecond timescale, corresponding to recombination across a *T*-dependent SC gap $\Delta_s$ and a *T*-independent „pseudogap" $\Delta_p$ respectively, where the magnitudes of the two gaps are doping-dependent (Figure 7).

The observation of two relaxation times at short timescales, one associated with the pseudogap $\Delta_p$, and the other with a *T*-dependent superconducting (or collective) gap by time-resolved spectroscopy in *all* cuprates so far (Figure 7), implies that the response is intrinsic and a universal at the fundamental level. The observation of a clear 2-component response in $La_{2-x}Sr_xCuO_4$ at all doping levels excludes the possibility that the two responses come from spatially separate parts in the crystal, such as charge-reservoir layers and $CuO_2$ planes respectively. Thus the 2-component response can be attributed to the $CuO_2$ planes, and implies a 2-component ground state, which – considering the resolution of the optical time-resolved experiments - can only exist on a mesoscopic or microscopic length scale.

An interesting feature of the systematic study shown in Figure 7 is the fact that the pseudogap $\Delta_p$ measured on the femtosecond timescale appears to be of similar magnitude in LSCO, YBCO and Hg-1223, in spite of the fact that $T_c$ differs by nearly a factor of 4. The temperature-dependent gap $\Delta_s(0)$ on the other hand, is closely related to $T_c$, giving a nearly constant *gap ratio* at optimum doping $2\Delta_s^{opt}(0)/kT_c = 9.4$ (LSCO), 10.1 (YBCO) and 12 (Hg-1223). We have recently shown that the relation between two gaps can be determined by the percolative threshold for phase coherence at finite temperature[57], or more concisely, given the characteristic energy scale $\Delta_p$ for the formation of pairs, then $T_c$ (and whence $\Delta_s(0)$) is determined by the geometrical size of the pairs or objects filling the plane *and* their number, which depends on temperature, as we shall see later.

## Model description of the dynamically inhomogeneous state

Very soon after the discovery of superconductivity in cuprates, the possible existence of electronic inhomogeneity was recognized by Gorkov and Sokol[42], while the possibility of spin stripes was discussed by Zannen and Gunnarson[43], Emery and Kivelson[44] and others. But it has taken quite some time to show experimentally that these inhomogeneities may be relevant to understanding the normal state and superconductvity[1] and obtain specific details which are necessary for a concise model to be proposed.

Here we approach the problem of the cuprates from the viewpoint that bosonic pairs are just a special kind of doping-induced impurity state. For specific reasons the materials prefer to form pairs rather than single polarons: competition between highly anisotropic elastic strain, exchange energy and Coulomb repulsion $V_C$ leads (at low doping) to the formation of nanoscale pairs in preference over single polarons.

The doped holes don't have sufficient binding energy $E_B$ to form bound states at high temperatures $kT > E_B$ and behave as a Fermion gas. For $kT^* < E_B$, a new structural phase starts to nucleate surrounding doped holes (i.e. polarons of well-defined lattice symmetry), leading to the coexistence of pairs and larger objects such as stripes, particularly as doping increases. In a more usual system exhibiting a structural phase transition, the length of these objects grows, showing critical behaviour, as the structural phase transition temperature $T_s$ is approached. However, in the cuprates, it appears that the doped holes which precipitate in the formation of a new phase also prevent long-range structural ordering, favoring the formation of a superconducting state instead. The small energy difference between pairs and clusters (stripes) leads to their co-existence in the pseudogap state below $T^*$.

In this section we shall concentrate on a microscopic description of the inhomogeneous state, starting with an examination of the symmetry of the interaction from inelastic neutron scattering experiments. We then describe an effective symmetry-invariant Hamiltonian and discuss its consequences for the appearance of an inhomogeneous state, and how this can lead to pre-formed pairs, stripes and superconductivity.

***The symmetry of the interactions and some consequences.***
The key to determining the symmetry of the interaction Hamiltonian for the system comes from the idea that the finite size coherence length $\xi_s$ defines the extent of the pair, which in turn means that the interaction between particles composing a pair is not at $k = 0$, but at finite $k = \pi/\xi_s$. Thus, any lattice or structural effect associated with the formation of pairs of size $\xi_s$ should result in a visible phonon anomaly in $k$-space at a wavevector corresponding to $k_0 = \pi/\xi_s$. Examination of the $k$-dependence of the phonon dispersion spectrum, measured by INS[17,18], or IXS[19] appears to strongly support this notion. Anomalies at long wavelengths ($k \to 0$) are not evident (or small). Rather, the largest anomalies appear at large $k$ (towards the zone boundary) in hole-doped cuprates[17,18] and slightly smaller $k$ in electron-doped cuprates[19], yet always consistent with $k_0 \approx \pi/\xi_s$. The point in $k$-space where an anomaly occurs also uniquely identifies the symmetry of the interaction which causes it. More precisely, the observation of a phonon anomaly at a particular point in $k$-space can be used to define the interaction between phonons, electrons (or holes) and spins, which is symmetry-invariant under the operations of the appropriate symmetry group of $k$ at that point in the Brillouin zone. The phonon dispersion anomalies for $La_{2-x}Sr_xCuO_4$ are in the $(\xi,0,0)$, or $\Gamma \to M$ direction [18]. The group of $k$ in this direction has $C_{2v}$ symmetry[45]. Hence an interaction which causes an anomaly at this point in k-space will lead to a distortion of the lattice whose symmetry is reduced from tetragonal symmetry ($D_{4h}$) within a volume of size $l = \pi/k_0$ [45]. The interaction at $k_0$ thus creates mesoscopic areas (or objects), whose symmetry is reduced (ultimately $C_{2v}$ - once larger objects corresponding to a new phase are formed), while the undistorted surroundings have a higher symmetry ($D_{4h}$). Note that the symmetry of small objects whose diameter is of the order of a unit cell cannot be easily defined, and strictly speaking, the proper symmetry group can be used only when larger, meso-scale objects are formed (by analogy to the symmetry transformation taking place at a structural phase transition).

Knowing the symmetry of the interaction enables one to write a Hamiltonian $H_{MJT}$ with the correct symmetry properties[45], which acts on a length scale defined by $l_0$. This effective interaction acts on a scale given by the length-scale of the inhomogeneities, and can be written in either real space or $k$-space. The real-space formulation is given as[45,46]:

$$H_{JT} = \sum_{\mathbf{n},\mathbf{l},s} \Bigg( \sigma_{0,\mathbf{l}}\{(n_x^2 + n_y^2)g_0(\mathbf{n})\}(b^\dagger_{\mathbf{l}+\mathbf{n}} + b_{\mathbf{l}+\mathbf{n}}) + \\
\sigma_{3,\mathbf{l}}\{(n_x^2 - n_y^2)g_3(\mathbf{n})\}(b^\dagger_{\mathbf{l}+\mathbf{n}} + b_{\mathbf{l}+\mathbf{n}}) + \\
\sigma_{1,\mathbf{l}}\{n_x n_y g_1(\mathbf{n})\}(b^\dagger_{\mathbf{l}+\mathbf{n}} + b_{\mathbf{l}+\mathbf{n}}) + \\
\sigma_{2,\mathbf{l}}S_{z,\mathbf{l}}\{(n_x^2 + n_y^2)g_2(\mathbf{n})\}(b^\dagger_{\mathbf{l}+\mathbf{n}} + b_{\mathbf{l}+\mathbf{n}}) \Bigg).$$

(4)

where $b$ and $b\dagger$ represent phonon operators and $S_z$ is the $z$ component of the local spin. $\sigma_i$ are the Pauli matrices representing the degenerate electronic states. The first term describes the coupling between electrons in non-degenerate states and is isotropic. The next two terms describe coupling between electrons in degenerate electronic states and the lattice of $x^2-y^2$ and $xy$ symmetry respectively, while the last term described the coupling of the $z$-component of spin $S_z$ to the electronic states. (All other couplings are forbidden to first order). The coupling functions $g_i(\mathbf{n})$ are given by:

$$g_i(\mathbf{n})=g_i\frac{\exp[-(a/l)\sqrt{n_x^2+n_y^2}]}{n_x^2+n_y^2} \quad (5)$$

Where $\mathbf{n} = (n_x, n_y)$ with $n_x, n_y \geq 1$, and $g_i$ are the coupling constants. The function $g_i(\mathbf{n})$ describes the spatial dependence of the effective interaction, as shown in Figure 13. It is an effective interaction which takes into account the isotropic Coulomb repulsion $V_c(r)$ and the attraction due to an anisotropic lattice strain, conforming to the crystal symmetry properties. A similar interaction can be written in $k$-space[46], except that $g(k)$ then describes the effective interaction with a peak at $k_0=2\pi/l$. The electron-lattice interaction described by the first term in (4) is isotropic. In principle all phonons which have the appropriate $\tau_1$ symmetry can couple to the electronic states. The second and third terms describe an interaction with anisotropic *d*-like deformations of $xy$ and $x^2-y^2$ symmetry respectively, and the last term describes the spin interactions with the out-of-plane component $S_z$. (In-plane components are not allowed by symmetry). One of the main conclusions from this analysis[45,46] is that non-degenerate electronic levels coupling to phonons and spins can only give rise to a symmetric (*s*-wave) deformation, while coupling to *doubly degenerate* electronic states (of $E_g$ symmetry in the high symmetry group) can give rise also to an anisotropic *d*-wave like interaction in addition to the symmetric one. Because the interaction is between electrons in degenerate states and phonons and spins, the latter may be viewed as a finite-*k* or *mesoscopic* Jahn-Teller effect (to distinguish it from the more standard single-ion JT effect, or cooperative JT effect – which leads to long range order).

An important feature of the effective interaction (4) is not only that it is highly anisotropic, but also that it peaks at finite *k* (or at finite range $l_0$). We have justified this on the basis of experimental observations, but we can also justify it theoretically if we consider the two most relevant fundamental interactions acting on doped holes, namely (highly anisotropic) elastic strain and an (isotropic) Coulomb repulsion. The total *V* is clearly highly anisotropic, and has a minimum at a distance *l*, depending on the difference between $V_c$ and $V_s$. The interaction (5) is a simplification of the more general form in Figure 13. The anisotropy arises simply as a consequence of orbital symmetry of the constituent atoms (primarily Cu *d*-orbitals and O *p*-orbitals) and the crystal symmetry.

The state which $H_{JT}$ (4) implies is dynamically inhomogeneous[46], where distorted and undistorted regions coexist (Figure 14) with a inhomogeneity on a scale defined by $k_0$ or $\xi_s$, but have different symmetry and different energies. In real space, the interaction $H_{JT}$ describes polaron-like objects of size $l_0 \approx \pi/k_0$, whose symmetry is reduced compared to the bulk of the crystal.

A very important question for superconductivity is how doped holes might order within such a model. Taking into account that the particles are charged and including the repulsive Coulomb interaction in its general form, we have shown that (4) can be reduced to a lattice gas model[47]. Taking only the $x^2$-$y^2$ term in (4), the interaction becomes:

$$H = \sum_{i,j}\left[-V_l(i-j)S_i^z S_j^z + V_C(i-j)Q_i Q_j\right]$$

(6)

where $S_i^z$ is an operator describing the anisotropic lattice interaction (4), whose value is $S_i^z=1$ for the state ($n_{i,1}=1, n_{i,2}=0$), $S_i^z=-1$ for ($n_{i,1}=0, n_{i,2}=1$) and $S_i^z=0$ for ($n_{i,1}=0, n_{i,2}=0$). $Q_i=(S_i^z)^2$. Here $n_{i,\alpha}$ are the occupation numbers of two-fold degenerate states $\alpha = 1,2$ at site $i$. The Coulomb interaction is formulated as $V_C(\mathbf{m})=e^2/\varepsilon_0 a(m_x^2+m_y^2)^{1/2}$, where $e$ is the electric charge, $a$ is the lattice constant and $\varepsilon_0$ is the dielectric constant and $\mathbf{m}=(m_x,m_y)$ is a vector. The attractive lattice interaction is written as:

$$V_l(\mathbf{s}) = (1/\omega)\sum_m \left(m_x^2 - m_y^2\right)\left\{(s_x + m_x)^2 - (s_y + m_y)^2\right\} g(\mathbf{m}) g(\mathbf{s}+\mathbf{m})$$

(7)

The interaction (7) now describes the ordering of JT polarons on a lattice in an *x-y* plane in the presence of competing Coulomb interaction $V_C$ and short range anisotropic attraction, where $g(\mathbf{m})$ has been defined in Eq. (5) and $\mathbf{s} = (s_x,s_y)$ is a vector. The short-range attraction is generated by the interaction of electrons with optical phonons, whose range is determined by the dispersion of optical phonons.

In principle, the objects created by the interaction can be single polarons, bipolarons, stripes etc., the stability of which is determined primarily by the balance between elastic energy $V_l$ and Coulomb repulsion $V_C$ on the mesoscopic scale. Monte-Carlo studies have shown[47] that the mode (5) can lead to the formation of both pairs and mesoscopic objects such as stripes and clusters, depending only on the ratio of parameters $V_C/V_l$ and the doping level. Both *xy*- and $x^2$-$y^2$ - symmetry stripes are allowed by symmetry, corresponding to "diagonal" stripes or stripes parallel to the bond axes respectively. Indeed, STM measurements strongly suggest that these exist, but may have different energies[11]. However, for the majority of experiments, the energy difference separating the distorted and undistorted regions can be associated with the pseudogap $\Delta_p$, and the state can be very effectively described in terms of a two-level system[6, 8,45,46,48] of spin singlets (S = 0) in the ground state, and unpaired (S = 1/2) spins in the excited state. The existence of *pairs* would imply the existence of spin singlets (S = 0) in the ground state, and unpaired (S = 1/2) spins in the excited state. Conversely, the presence of single isolated polarons would give rise to a substantial Curie susceptibility due to unpaired localised spins. Moreover the spin susceptibility at any give $T$ increases with doping. Hence - as already discussed - the existence of single polarons in the ground state is categorically ruled out by magnetic measurements, while static susceptibility measurements[8,20,48], and NMR Knight shift data[20] (Figure 15), support the proposed model description of spin singlets in the ground state and S=1/2 Fermions in the excited state. The magnetic response is thus quantitatively consistent with the existence of pre-formed Jahn-Teller *bi*polaron pairs in the ground state[8,45,46,49]. The existence of stripes of different size and direction is then described by intra-gap states. Note that the same two-level system has also been used to fit the temperature-dependence of the femtosecond-timescale time-resolved experiments, while the intra-gap relaxation proceeds on a slower, nanosecond timescale and exhibits glass-like stretch-exponential relaxation, consistent with stripes as intra-gap states.

### *Evidence of the symmetry-breaking interaction: loss of inversion symmetry.*

The distorted objects have reduced symmetry, so we expect to observe manifestations of symmetry breaking as pairing distortions start to occur. Evidence for local inversion symmetry breaking below the pseudogap temperature is plentiful both in Raman[50],[51] and infrared spectroscopy[52]. For example, the $T$-dependence of broken-symmetry Raman modes rather convincingly follow 2-level-system behaviour[50] (see

Figure 16). Rather convincingly, there is also a large amount of evidence for the existence of a spontaneous polarization in cuprates over a wide range of doping[53], with the observation of phenomena such as pyroelectricity and piezoelectricity[54], which imply the existence of on a non-centrosymmetric structure, such as $C_{2v}$ symmetry. Note that this inversion symmetry-breaking is a specific prediction of the model (4), and to our knowledge, there is no other theoretical model currently available which predicts these symmetry breaking phenomena in the cuprates.

### *Stripes and pairs: overcrowded state*

At high doping density, the distorted regions might aggregate to form longer stripes or clusters, with the same symmetry properties defined by $H_{JT}$ (henceforth we generically use the term stripes to signify any aggregated object larger than a pair). A comprehensive discussion of superconductivity involves the formation of a phase-coherent state within such a system of co-existing pairs, "stripes" and – at finite temperature – Fermions in the excited state. The *T*-dependence of the length-scale $l_0$ measured by QP recombination experiments (Figure 10) is an indication that stripes and clusters form at low temperatures.

However, it is not immediately obvious whether stripes should have significantly lower energy than single bipolarons. To be relevant, the energy difference between bipolarons and stripes should be at least on a scale of $kT_c$, giving a more complicated energy level structure, e.g. a 3 level system. One answer to the question whether there is an additional energy scale *for different types of stripes or clusters* comes from experiments.

One indication that the state has more than one component has come from time-resolved experiments on the femtosecond timescale, where very early, there was evidence for the co-existence of metallic and localised states[1,21,22].

A careful fit to the magnetic susceptibility data[8] also shows that a two-level description originally proposed by Alexandrov, Kabanov and Mott[56] is not completely sufficient. A more accurate fit to the susceptibility data can be obtained when a temperature-independent Pauli-like contribution is included, yet only one energy scale ($\Delta_p$) is used to fit the whole data set, suggesting that the excitation energies for bipolarons and stripes are indistinguishable, or very close in energy. Note that when stripes are included[8] overall there are two Fermionic contributions to the susceptibility in the mixed state of pairs and stripes: (i) At high temperatures (above $T^*$), we have a Fermi gas, where the susceptibility is nearly $T$-independent, and there are neither bipolarons nor stripes present. (ii) At low temperatures, both bipolarons and stripes are present, the latter with a Pauli-like contribution to the susceptibility. This temperature-independent susceptibility can then be understood to come from metallic stripes[8]. (A similar conclusion regarding metallicity of the stripes was derived on the basis of theoretical arguments by Mihailovic and Kabanov[45].) Yet further indication for the existence of stripes or clusters with energy states within the pair-gap energy scale is the slow and glass-like dynamics of the intra-gap state relaxation[24] and recent ESR data show the existence of two components even in lightly doped $La_{2-x}Sr_xCuO_4$ ($0.01<x<0.06$)[55].

We can conclude that there appear to be strong experimental and theoretical arguments for the co-existence of bipolarons and larger objects, such as stripes in cuprates not only at optimum doping and in the overdoped state, but also in the underdoped state. Clearly pairs must exist in the ground state, otherwise at very low temperatures, the cuprates would not be superconducting.

## Superconductivity

There are two fundamentally different types of dynamics which need to be considered in the discussion of superconductivity in an inhomogeneous system:

1. In the first case, the inhomogeneity is simply thermally fluctuating according to the relevant statistics defined by the energy scale of the bound state $\Delta_p$ in relation to *kT*. In other words, charge-rich inhomogeneities (polarons or bipolarons) may form, disappear, and reform in different spatial locations, driven by thermal fluctuations. The inhomogeneities are still static in the sense that the objects have *no center-of-mass motion,* i.e. their kinetic energy is small compared to their binding energy. Nevertheless their presence fluctuates on a fast timescale $\tau \sim \hbar/\Delta_p$ (of the order of $10^{-14}$ s).

2. The other possibility is that the dynamic inhomogeneity is associated with the center of mass *motion* of bosons (bipolarons).

For some models of superconductivity (Bose-Einstein condensation for example) it is essential that the pairs are mobile[56] (case 2). For other models (such as percolative superconductivity[57]) motion of pairs is not essential. It is thus important to determine experimentally which of the two cases applies to the cuprates.

A common objection to superconductivity models based on Bose condensation of bipolarons in the cuprates is based on „overcrowding". In order to ensure that the interparticle distance is greater than the effective bipolaron radius $r_0$, the latter has to be smaller than the coherence length $\xi_s$, i.e. $r_0 \ll \xi_s$. In the cuprates this might be satisfied only in the dilute (strongly underdoped) case, while in the optimally doped and overdoped cuprates the model is probably not applicable.

An alternative scenario proposed by the present authors[57] considers bipolarons where the kinetic energy of the pairs plays no part, whereby a macroscopic superconducting state is formed by phase coherence percolation by Josephson coupling (i.e. pair tunnelling) between deformed regions, *including* coupling across metallic stripes. A schematic picture of a phase coherence percolation is shown in Figure 17. It was shown[57] that if each doped carrier created a distortion of size $\pi l_0^2$ (or $\pi \xi_s^2$, since $l_0 \approx \xi_s$ at $T_c$) in the CuO$_2$ plane of a cuprate superconductor, then at the percolation threshold, the doped carrier density corresponds to the onset of the superconducting state. The percolation threshold for 2D is at ½ area fraction in a bond-percolation model (which is a good approximation to the case in hand). In La$_{2-x}$Sr$_x$CuO$_4$, this corresponds to x = 0.06 at T = 0.

Doping beyond this point leads to filling beyond ½ effective 2D volume fraction, there are more stripes and fewer pairs, whence the percolation threshold between pairs is reached at lower temperature, in effect lowering $T_c$ from it would be if space were not filled with stripes. In this picture, stripes are detrimental to high $T_c$s. Higher $T_c$ appear to be limited by the topological necessity that overlapping pairs form (Fermionic) stripes at higher doping.

The superconducting gap $\Delta_s(T)$ appears as result of long-range order arising from Josephson coupling across stripes and pairs, leading to the characteristic mean-field like temperature dependence of $\Delta_s(T)$ clearly observed for QP recombination in the femtosecond spectroscopy experiments. Hence the simultaneous presence of $\Delta_p$ and $\Delta_s(T)$ in the time-resolved optical data can be understood. In this model, the ratio of $T_c$ to $\Delta_p$ is dependent on the effective geometrical size of the pair (which is in turn related to the coherence length), as discussed previously in this review and in Ref. 57. It should be mentioned here that if pairs can tunnel between inhomogeneities[57], it is not particularly important whether the inhomogeneities are dynamic or static, provided the inhomogeneities exist on a timescale which is longer than the pair tunneling time.

**Conclusions**

In this review we have presented an overview of the dynamical state of cuprate superconductors based on experimental observations on short timescales. The observed behaviour appears to be quite universal, defining the underlying physical mechanisms which lead to the appearance of superconductivity in these materials. We have also described a theoretical model to account for the dynamically inhomogeneous state, which quantitatively and self-consistently describes many of the general features of the observed dynamics on the femtosecond timescale as well as experiments on slower timescales, such as NMR[20] and magnetic susceptibility[8]. The model also predicts very specific symmetry breaking phenomena observed in Raman[50,51], infrared spectroscopy[52] and a class of ferroelectric phenomena[53,54] which are also observed in cuprate superconductors. Moreover, transport properties, particularly resistivity, have also been quantitatively described by model[58].

The emerging picture is one in which the inhomogeneity is coincident with singlet pairing, while the formation of a superconducting state at $T_c$ is governed by Josephson coupling between inhomogeneous regions[57]. We are not aware of any other approach which can quantitatively describe such a wide variety of phenomena within a single model.

We conclude by noting that formally, we have divided the description of superconductivity into three parts: (i) a microscopic model for formation of bipolaron pairs and stripes as the dominant form of inhomogeneity, considering primarily the competition of highly anisotropic elastic attraction and Coulomb repulsion $V(\mathbf{r})$ between particles, concisely taking the appropriate symmetry considerations into account, (ii) a set of 2 level systems describing the pseudogap behaviour of pairs and stripes, and (iii) the formation of superconducting state in an inhomogeneous medium via phase coherence percolation.

More experiments and modelling are necessary in order to clarify some of the details of mesoscopic ordering of carriers into stripes[59]. Although one should expect that a detailed examination of new experiments will reveal additional complexity, it seems that the proposed finite-wavevector Jahn-Teller model can describe the salient features very successfully.

Acknowledgments. I wish like to thank K.A.Muller, A.R.Bishop, J.Demsar, T.Mertelj, S.Billinge, T.Egami, D.Pavuna and I.Bozovic for helpful comments and discussions.

Figure captions

Figure 1. The magnitude of the pseudogap from time-resolved QP recombination experiments compared to the energy scale of the anomalies observed in neutron data in YBCO (shaded region) [18], connecting lattice anomalies with QP recombination dynamics.

Figure 2. A schematic diagram of the photoexcitation and subsequent relaxation processes in cuprates. The pump a) and probe laser pulses b) are separated by a variable time delay. All possible probe processes are shown in b). Whether the change is a photoinduced absorption or bleaching, depends on *which* optical processes (1 to 4) are dominant. If by coincidence all probe transitions are equal, there is no photoinduced signal (see paper by Dvorsek *et al.* in ref. 30).

Figure 3. a) Typical time-resolved optical response in YBCO. The logarithmic scale (insert) emphasises the existence of two fast recombination components below $T_c$. One component disappears above $T_c$. The time-decay behaviour is very similar in other cuprates such as b) $YBa_2Cu_4O_8$, where the two components have different sign, depending on probe polarisation, c) in $La_{2-x}Sr_xCuO_4$ and d) Hg-1223.[30]

Figure 4. The temperature dependence of the photoinduced signal amplitude in a) LaSrCuO and b) YBCO and c) Hg-1223. Two distinct responses are observed. One disappears at $T_c$, while the other is associated with the pseudogap $T^*$ and asymptotically falls above that temperature. This remarkable behaviour is universal in the cuprates[23,24,30].

Figure 5. A schematic drawing of the temperature dependence of the two gaps inferred from time-resolved QP recombination experiments on YBCO. The pseudogap $\Delta_p$ might signifies the bipolaron binding energy, while the $\Delta_s(T)$ is the temperature-dependent collective gap associated with the superconducting state.

Figure 6.  The QP lifetime as a function of temperature in a) YBCO, b) YBCO-124, c) LaSrCuO and d) Hg-1223[30]. Similar behaviour was also observed in Tl-1223, $Nd_{2-x}Ce_xCuO_4$ [33] and other cuprates.

Figure 7.  The magnitude of the pseudogap $\Delta_p$ and superconducting gap $\Delta_s(T)$ as a function of doping in $YBa_2Cu_3O_{7-\delta}$ and $Y_{1-x}Ca_xBa_2Cu_3O_{7-\delta}$, $La_{2-x}Sr_xCuO_4$. and Hg-1223. The remarkable feature of the data is that the energy gaps associated with pair recombination above $T_c$ are similar in magnitude, whereas the superconducting gaps are quite different, corresponding to the very different $T_c$s of the three materials.

Figure 8.  a) Recombination of 2 QPs proceeds with the emission of an optical phonon with energy $\hbar\omega_q > 2\Delta$. The re-absorption of the phonon breaks a pair. The repetition of this sequence leads to a Rothwarf-Taylor bottleneck.[35]

Figure 9.  Real-space representation of QP recombination showing the decay of emitted optical phonon into two acoustic phonons. When the acoustic phonons escape the bipolaron volume, they can no longer be re-absorbed to break pairs.

Figure 10.  Normalized length scale $l_0$ vs. $T/T_c$ determined from the phonon escape time $\tau_R$ in LSCO, Hg-1223 and NdCeCuO from ref. 33.

Figure 11.  Schematic diagram showing the initial relaxation of hot electrons excited by the pump laser pulses. The relaxation proceeds via carrier-carrier scattering and via phonon emission, whereby the particles encounter of the order of 50 scattering events before reaching the gap, losing memory of their initial momentum. Once they have reached the gap, they occupy all momentum states approximately equally.

Figure 12.  Recombination processes involving 2 QPs with $k_1$ and $k_2$ respectively (dashed arrows) and an emitted phonon (wavy line) for a) $T > T_c$ where the final state momentum is not necessarily zero, and b) the condensate $T$

< $T_c$ where the pair momentum is strictly zero (in the absence of current). The electronic structure is based on ARPES, whereby the shaded areas at the M point represent a large spectral density at an energy $\Delta$ below $E_F$.

Figure 13  The two most relevant potentials acting on particles in the cuprates are Coulomb ($1/r$) and strain (of the form $1/r^2$ in the continuum limit). The total potential has a shallow maximum, which gives rise to a rich energy landscape of stripes, checkerboards and bipolaron pairs. At short distances, Coulomb repulsion overcomes strain preventing on-site double occupancy. $V_{\text{eff}}$ describes an effective potential used to describe the short-range Jahn-Teller pairing interaction at finite $k$ [45,46,47].

Figure 14.  Real-space picture of a dynamically inhomogeneous state, depicting the $CuO_2$ plane of the cuprates. The shade of gray represents the sign of the effective interaction for the case of $x^2$-$y^2$ symmetry (second term in the Hamiltonian (4)).  Pairs and stripes are shown.

Figure 15.  The NMR knight shift as a function of temperature follows the behaviour expected for a 2-level system of bipolarons (S=0) and excited state (S=1/2) Fermions. [20].

Figure 16.  a) The temperature-dependence of the ground state and excited state poplation in a 2-level system with a gap $\Delta$ for YBCO. b) The T-dependence of the anomalous phonon intensity measured by INS[18]. c) The temperature-dependence of Raman scattering intensity for modes forbidden in the centro-symmetric high-temperature structure.  The broken-symmetry modes can be assigned to infrared modes, which become Raman-active when inversion symmetry is broken by the formation of pairs or stripes. The two samples correspond to the optimally doped YBCO ($T_c$=90K) and underdoped YBCO ($T_c$=84K) [46].

Figure 17.  Macroscopic phase coherence by Josephson percolation across pairs and stripes. The white areas depict charge density. The line indicates a phase-doherent percolation path.

# Figure 1

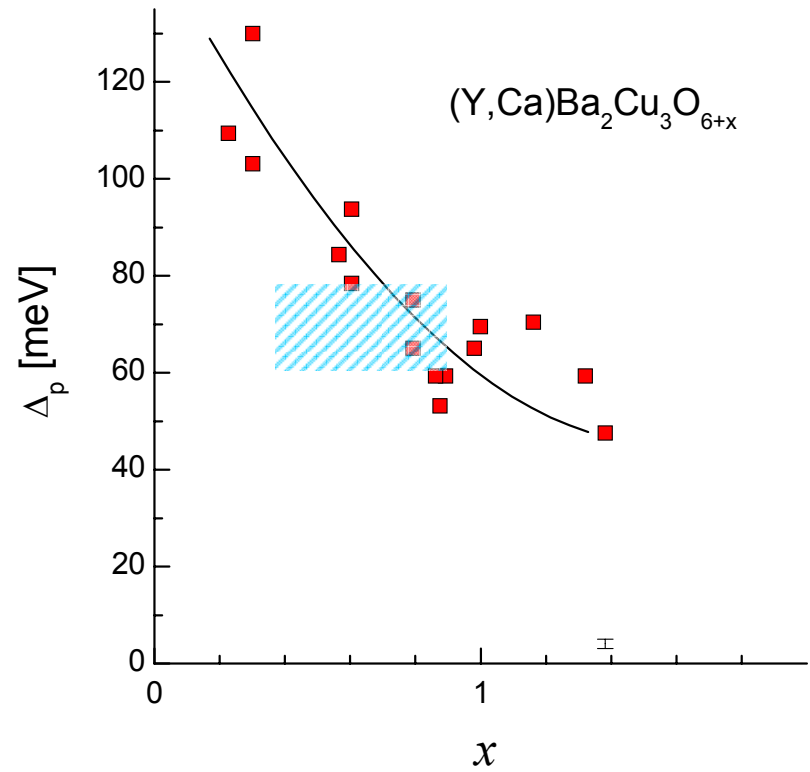

# Figure 2

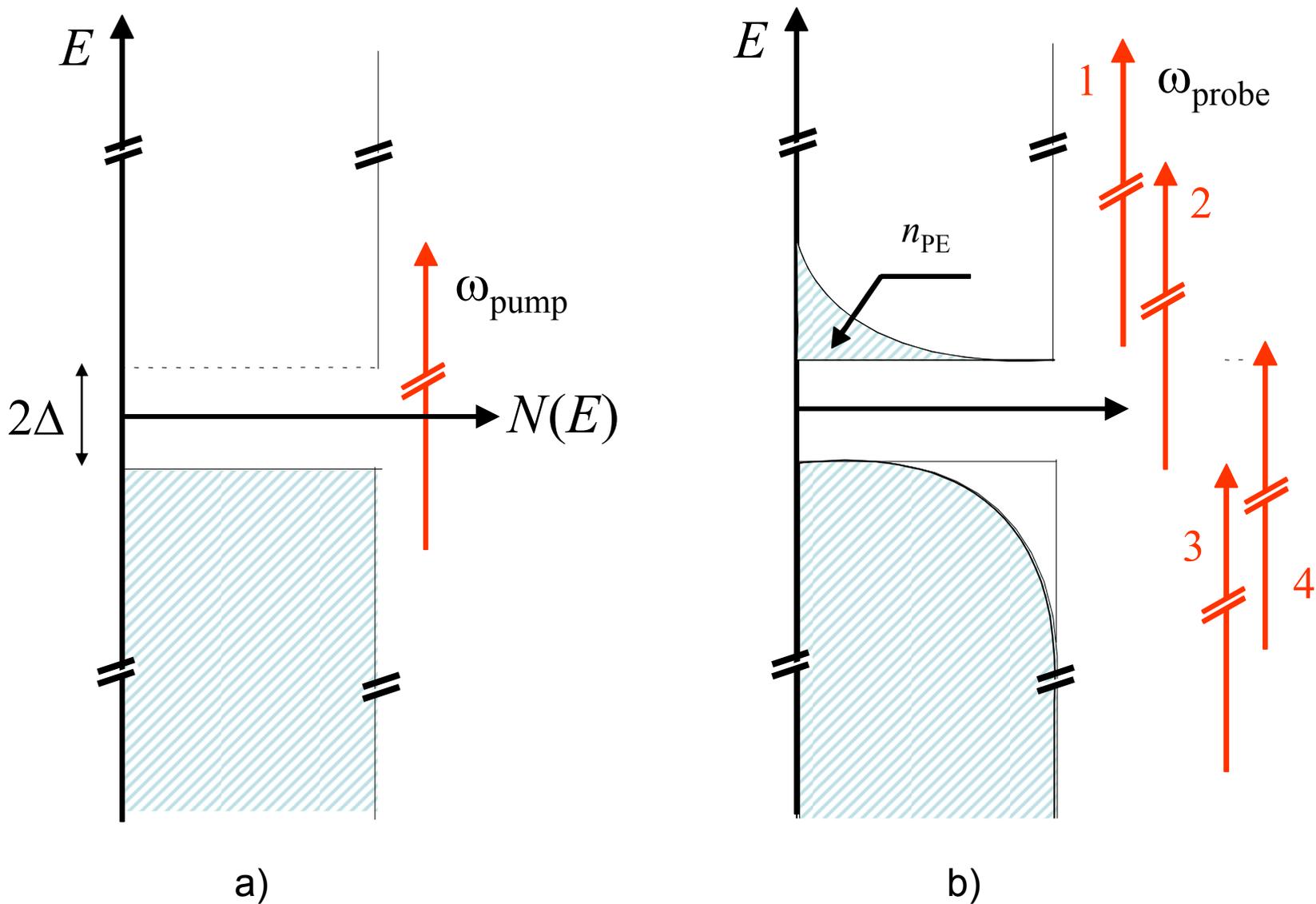

a)

b)

# Figure 3

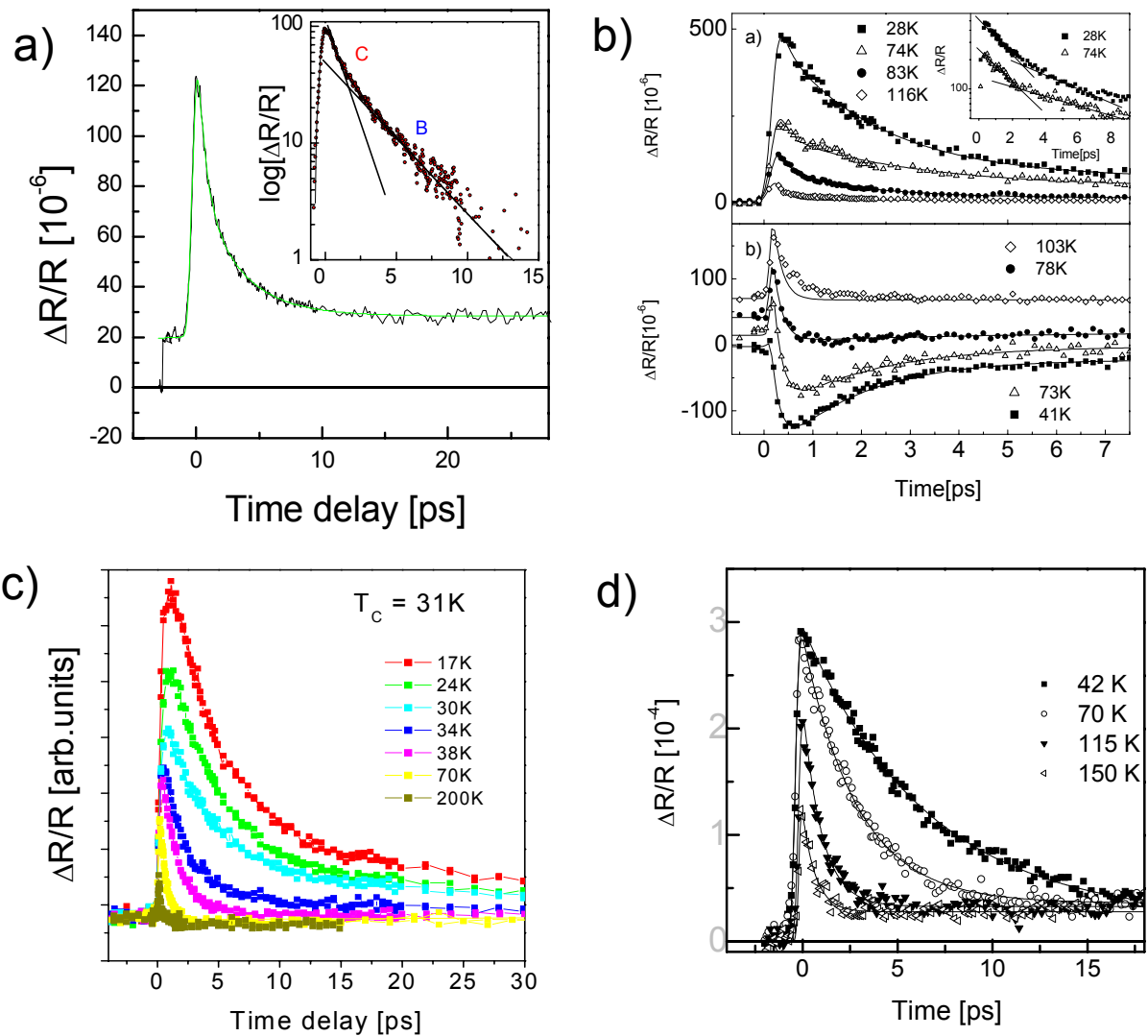

# Figure 4

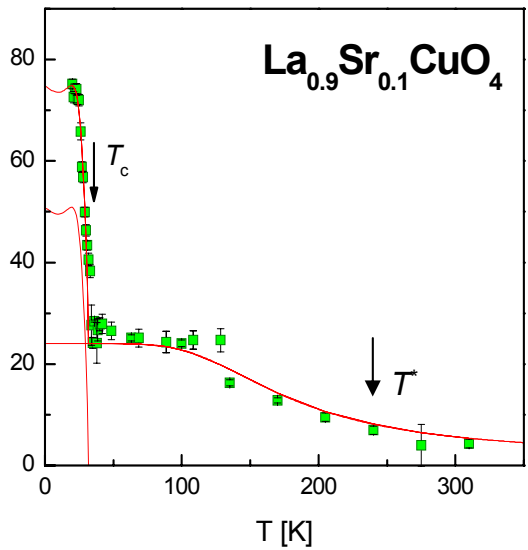
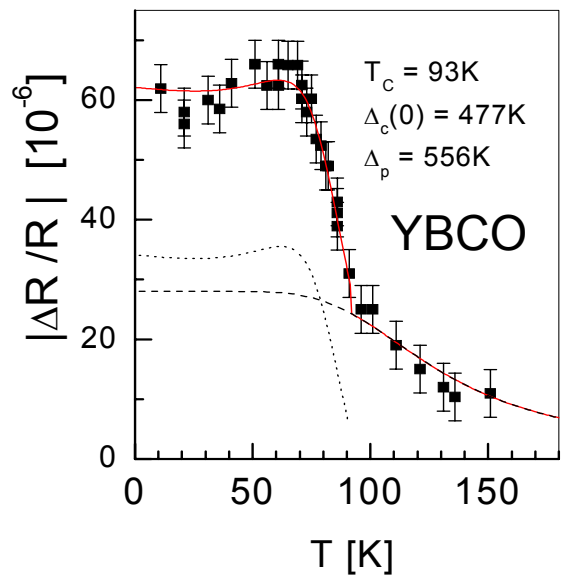
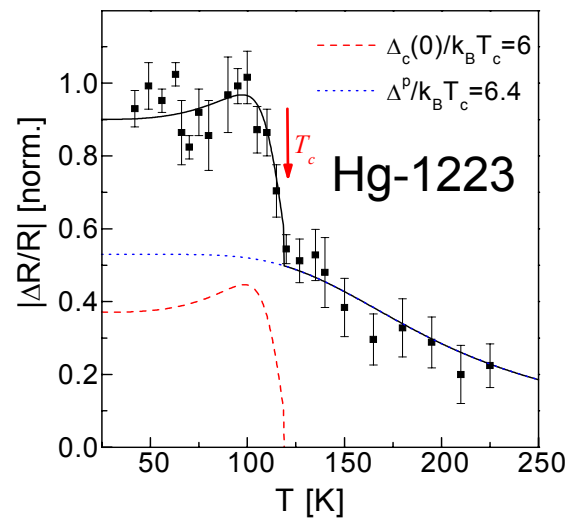

# Figure 5

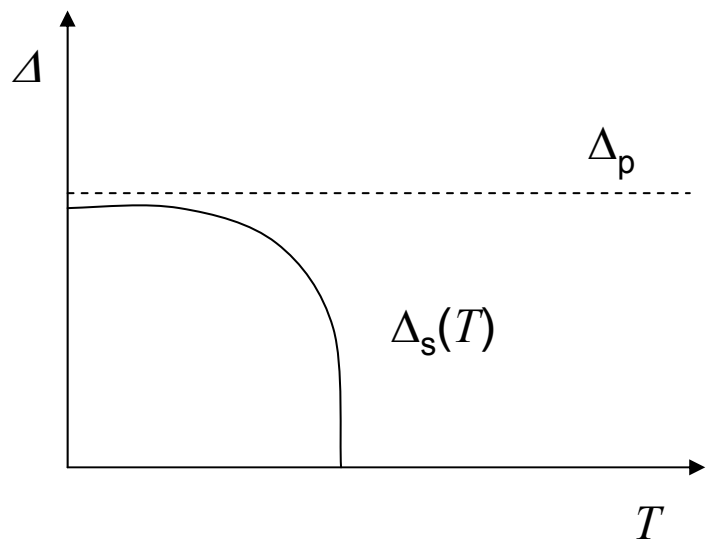 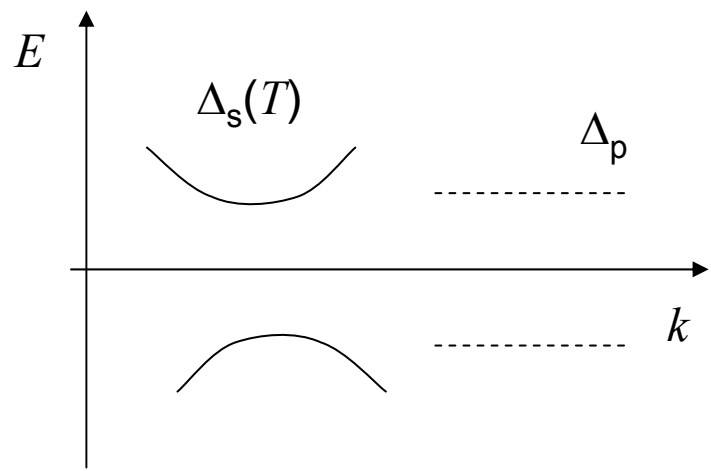

a)  b)

# Figure 6

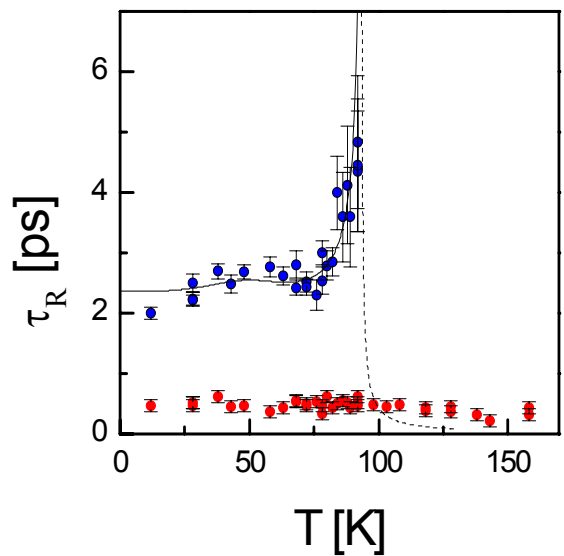
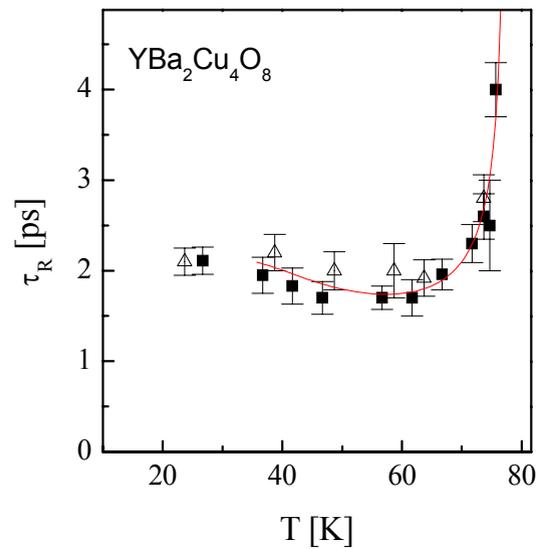
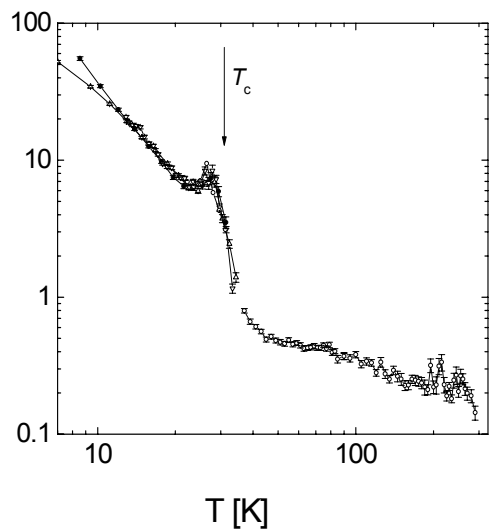
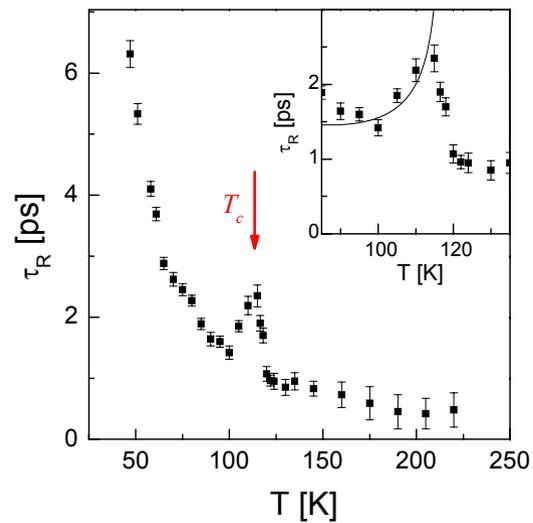

# Figure 7

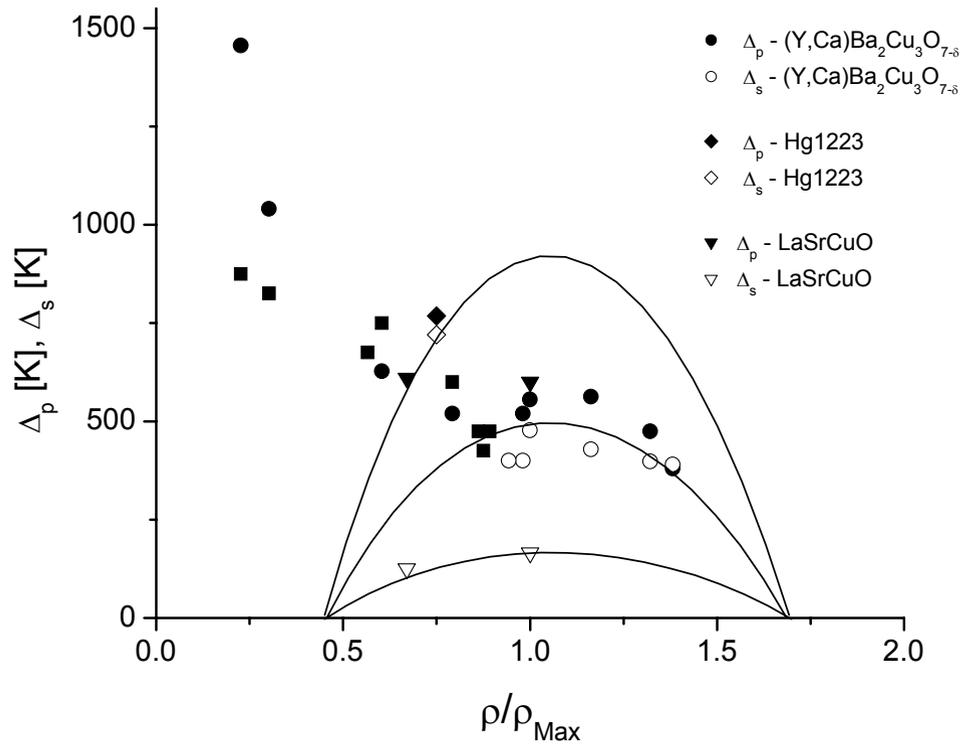

# Figure 8

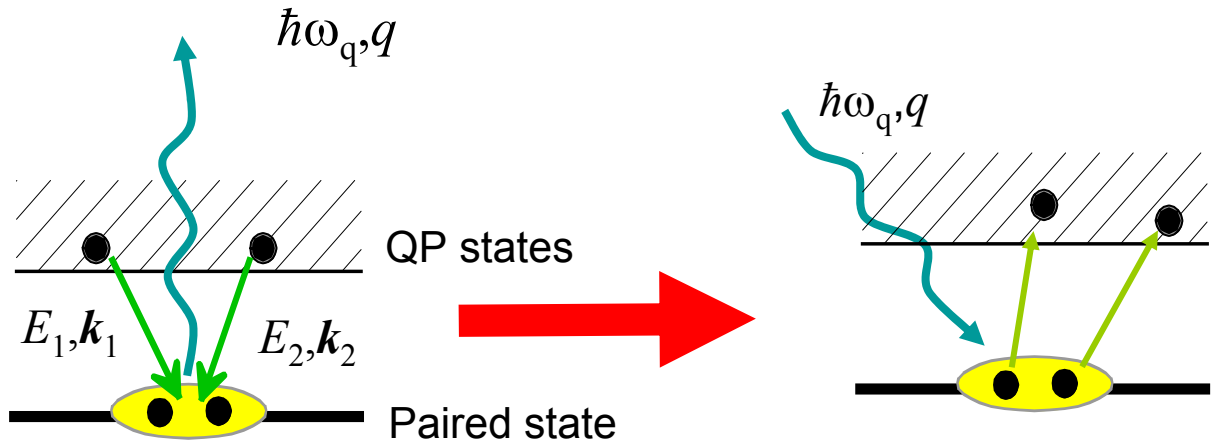

1. Emission of a phonon with $\hbar\omega_{phonon} > 2\Delta$

2. Re-absorption of a phonon with $\hbar\omega_{phonon} > 2\Delta$

# Figure 9

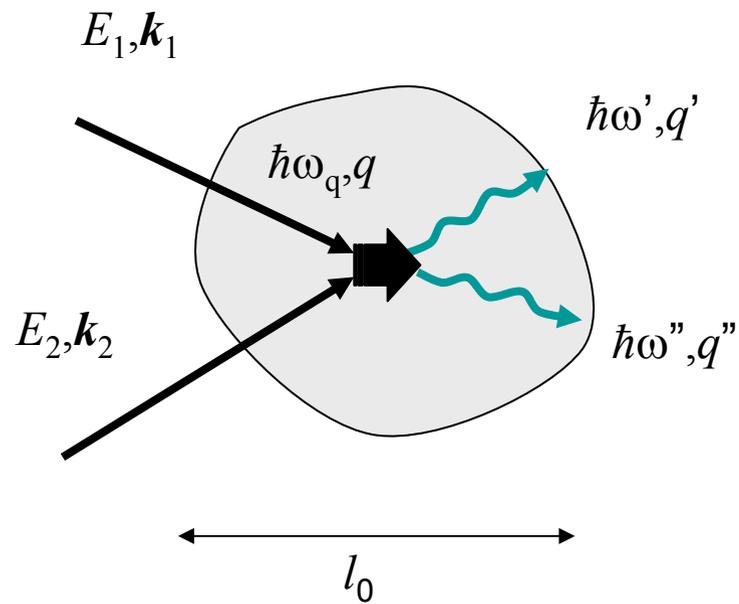

# Figure 10

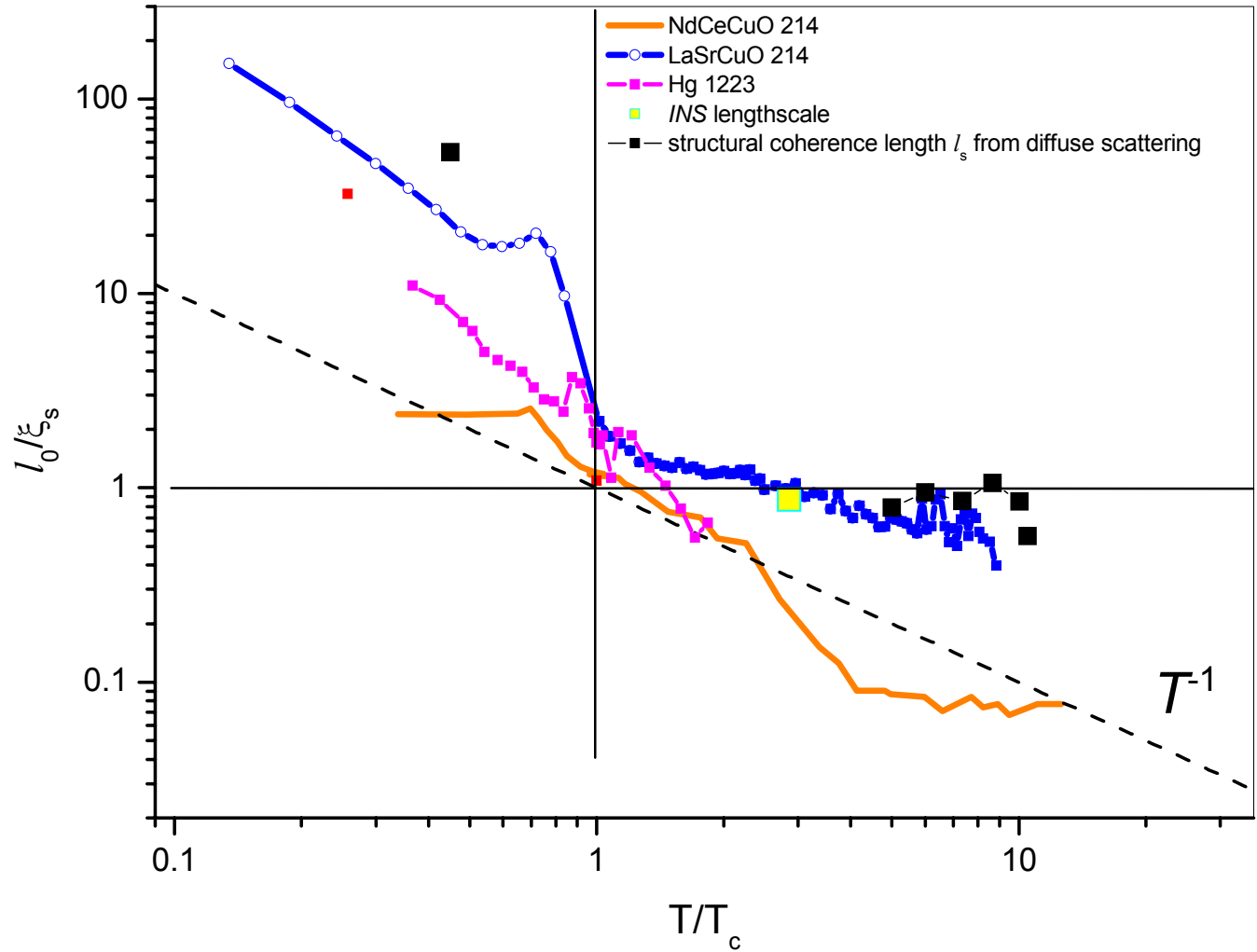

# Figure 11

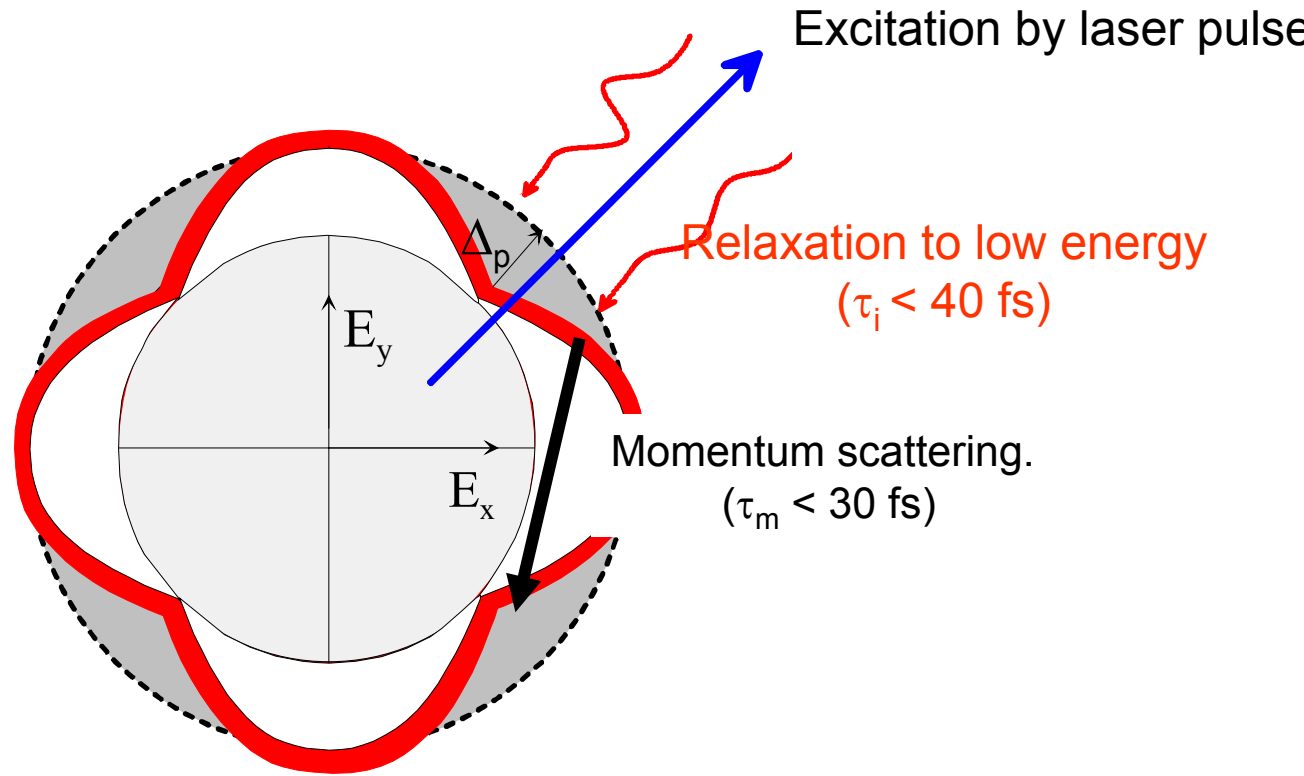

# Figure 12

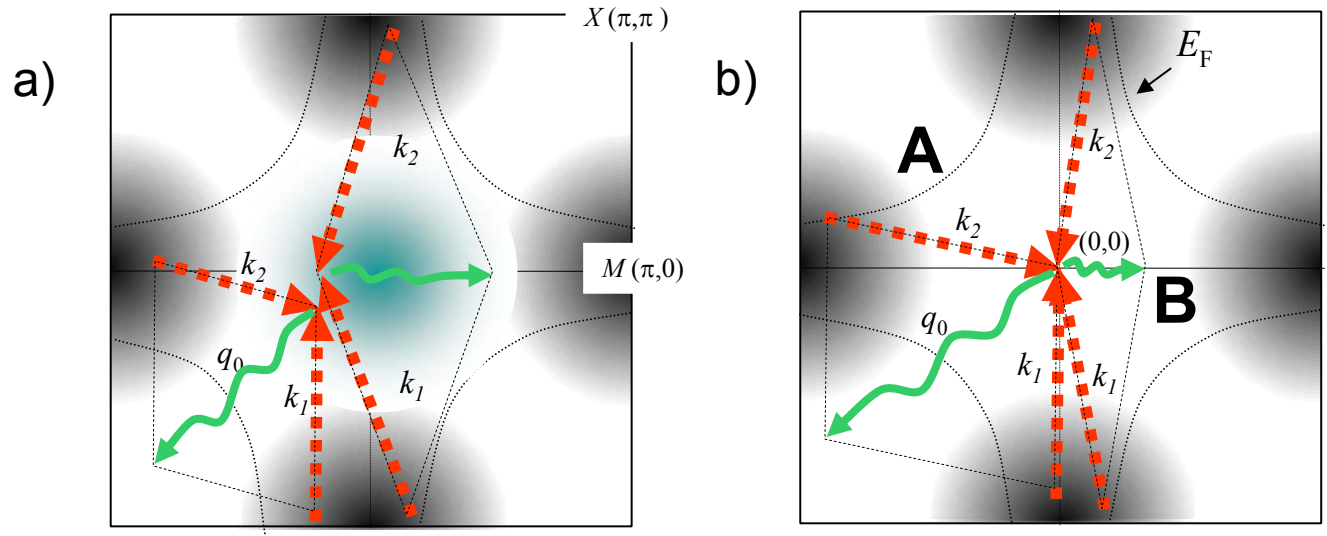

# Figure 13

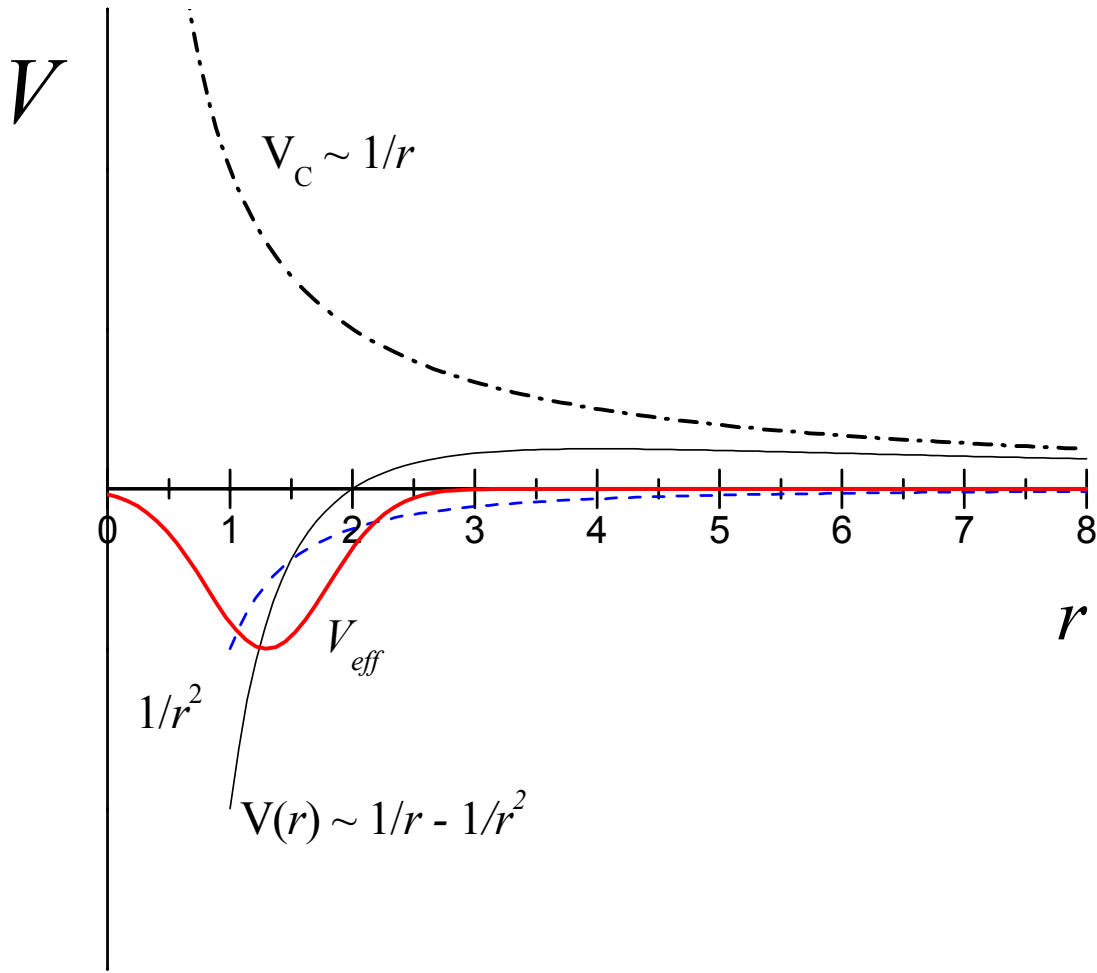

# Figure

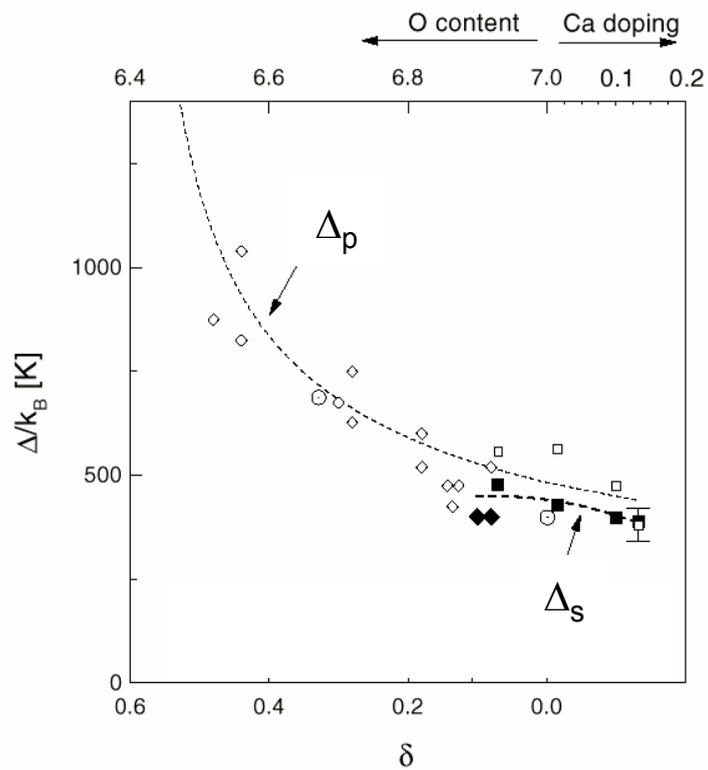 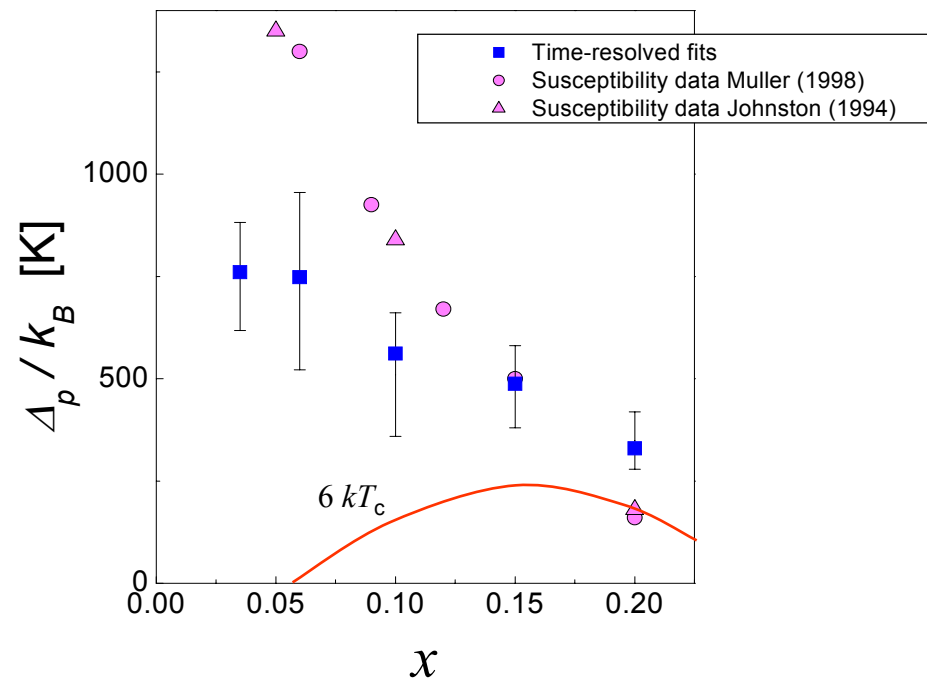

a)                  b)

# Figure 14

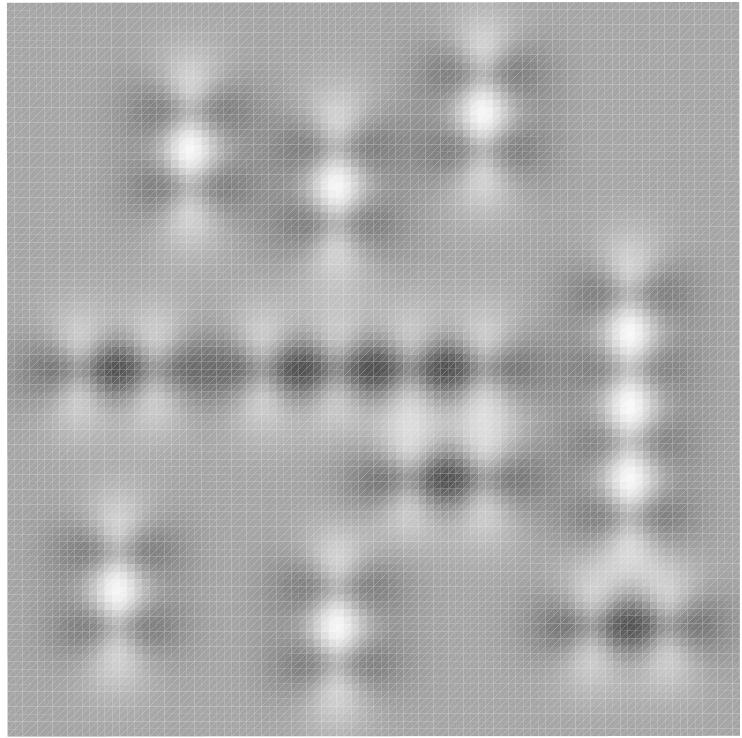

# Figure 15

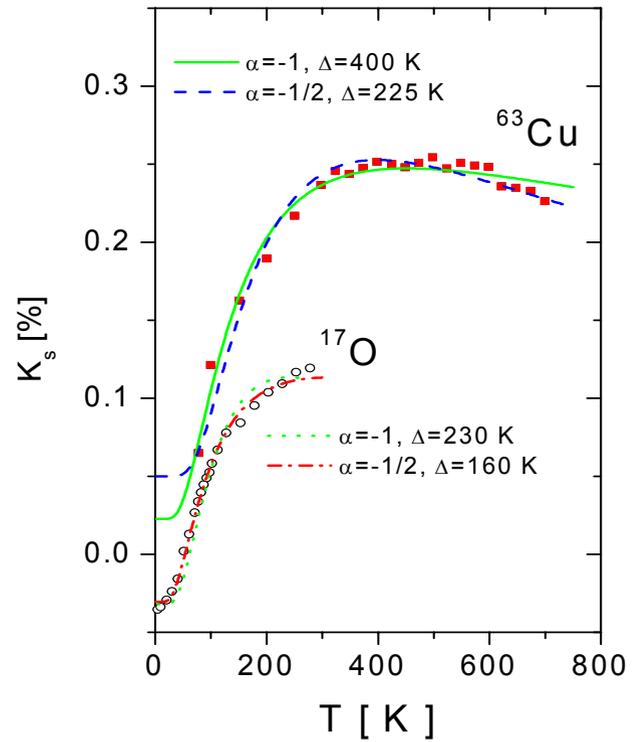

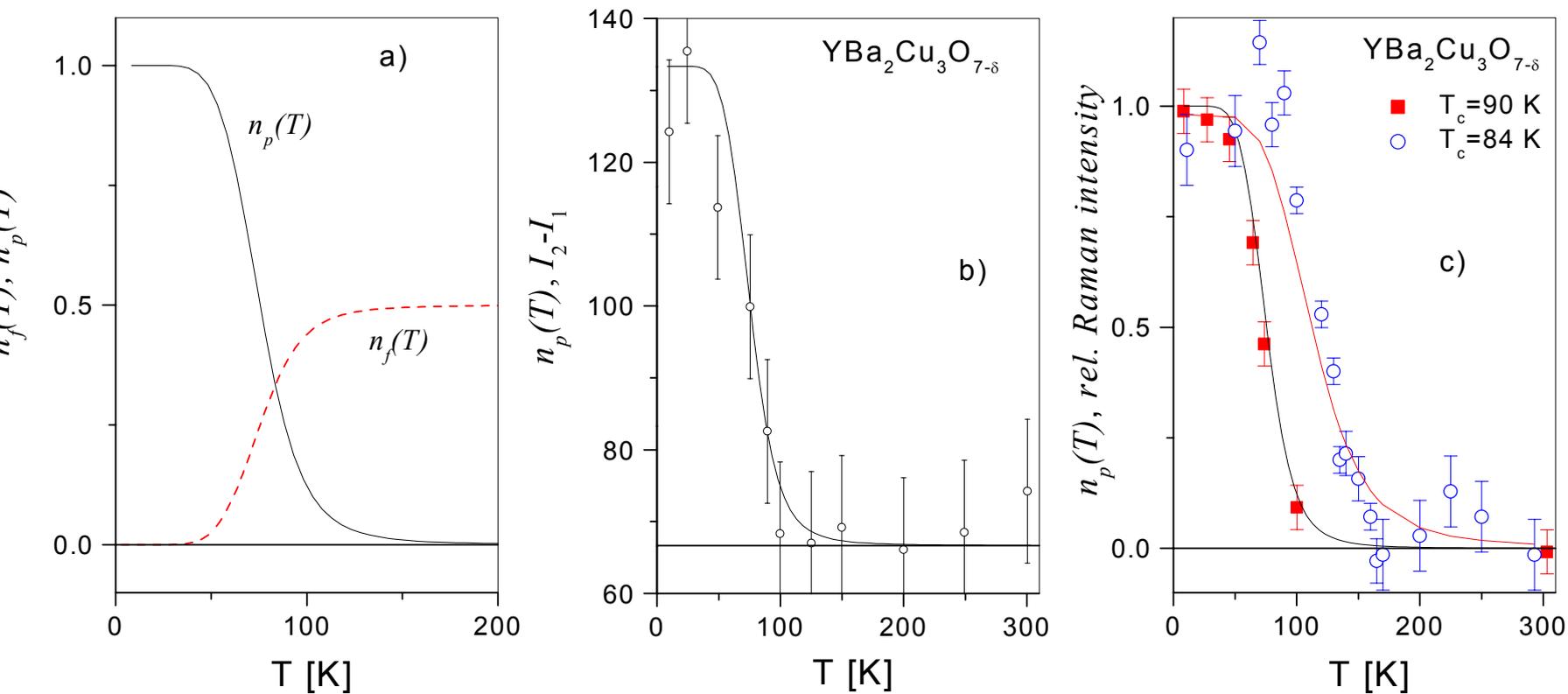

# Figure 17

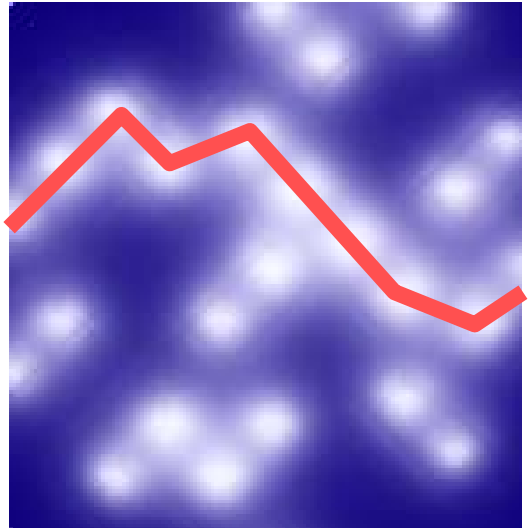